\documentclass[
aps,pra,
reprint,
superscriptaddress,
showpacs,
preprintnumbers,
amsmath,amssymb,
]{revtex4-2}

\usepackage{graphicx}
\usepackage{dcolumn}
\usepackage{bm}
\usepackage{xcolor}
\usepackage[colorlinks=true,
citecolor=blue,
urlcolor=blue,
linkcolor=blue,
pdfborder={0 0 0}]{hyperref}

\newcommand{\ket}[1]{|#1\rangle}
\newcommand{\bra}[1]{\langle #1|}

\newcommand{\fH}{\hat{\mathcal{H}}}

\newcommand{\ii}{\mathrm{i}}
\newcommand{\ee}{\mathrm{e}}
\newcommand{\dd}{\mathop{}\!\mathrm{d}}
\newcommand{\kB}{k_{\mathrm{B}}}

\begin{document}
	
	\title{A Universal Topological Platform for Nonreciprocal Spin-Photon Interface in Solid-State Quantum Networks}
	
	\author{Fang-Yu Hong}
	\email{honghfy@163.com}
	\affiliation{Zhejiang Key Laboratory of Quantum State Control and Optical Field Manipulation, Department of Physics, Zhejiang Sci-Tech University, Hangzhou, Zhejiang 310018, China}
	
	\date{\today}
	
	\begin{abstract}
		A fundamental obstacle to scalable solid-state quantum networks is the lack of a universal interface that simultaneously provides strong light-matter coupling, deterministic nonreciprocal photon routing, and efficient extraction from deeply subwavelength emitters. Here we propose and theoretically validate a plasmonic platform that overcomes these challenges by leveraging the unique properties of a Tomonaga--Luttinger liquid (TLL) formed in a single-walled carbon nanotube (SWCNT) microtoroid. We demonstrate that the TLL's collective bosonic excitations are kinematically protected against backscattering by a large valley-momentum mismatch, guaranteeing robust chiral spin-momentum locking that is unattainable in conventional dielectric cavities. This 1D kinematic protection enables deterministic routing of circularly polarized photons from a quantum emitter---here exemplified by a nitrogen-vacancy (NV) center---into distinct clockwise and counterclockwise propagation channels. By deterministically aligning the emitter's symmetry axis, parasitic $\pi$ transitions are strictly forbidden by geometry. Furthermore, we show that residual atomic-scale backscattering can be suppressed to the $\sim 100\,\mathrm{Hz}$ level via electrostatic gating and annealing. To overcome the severe mode mismatch between the CNT plasmon and the fiber, we introduce a graded plasmonic-photonic mode converter that adiabatically transforms the deeply confined CNT plasmon into a low-loss dielectric waveguide mode, providing a practical path to near-unity extraction efficiency. Using a tripod-STIRAP scheme, we show that the system can generate high-fidelity, magnetically tunable spin-photon entanglement. Our analysis confirms operation deep in the strong-coupling regime, with cooperativities \(C > 100\) and chiral contrast exceeding \(20\)\,dB. This architecture is fundamentally wavelength-agnostic and compatible with any solid-state emitter. It thereby establishes a scalable blueprint for robust, nonreciprocal quantum nodes that bridge the gap between individual spin qubits and a global quantum internet.
	\end{abstract}
	
	
	\maketitle
	
	\section{Introduction}
	
	The realization of a global quantum internet hinges on the development of efficient, scalable matter--photon interfaces~\cite{Kimble2008,Wehner2018,Briegel1998,Duan2001,Sangouard2011,Northup2014}. Solid-state spin qubits, such as the nitrogen-vacancy (NV) center in diamond, are prime candidates for such nodes due to their excellent spin coherence and optical accessibility~\cite{Awschalom2018,Doherty2013,Jelezko2006,Childress2006,Togan2010}. Despite remarkable progress in coupling these emitters to nanophotonic structures and demonstrating remote entanglement~\cite{Reiserer2015,Lodahl2015,Bernien2013,Hensen2015,Humphreys2018,Pompili2021,Sipahigil2016,Bhaskar2020}, a universal interface that simultaneously satisfies three critical requirements has remained elusive: (i) operation in the strong-coupling regime to enable deterministic spin-photon entanglement; (ii) intrinsic nonreciprocity for directional, low-loss photon routing without external magnetic fields or circulators; and (iii) near-unity extraction efficiency from the emitter's nanoscale environment into a standard optical fiber network.
	
	Conventional dielectric microcavities and waveguides, while offering high quality factors, are fundamentally constrained in achieving these goals simultaneously. First, their mode volumes are diffraction-limited, restricting achievable coupling strengths unless emitters are precisely positioned within nanoscale hot-spots~\cite{Vahala2003,Aspelmeyer2014,Mabuchi2002,Haroche2006,Burek2014,Riedrich-Moller2014}. More critically, they are inherently susceptible to backscattering from inevitable fabrication imperfections, which mixes counterpropagating modes and destroys the chiral routing fidelity essential for nonreciprocal operation~\cite{Hu2014,Fong2021,Mazzei2007,Kippenberg2002}. While extensive efforts have been made to mitigate these effects through optimized fabrication and chiral light-matter interaction designs~\cite{Petersen2014,Mitsch2014,Sollner2015,Coles2016,Lodahl2017,Bliokh2015,Aiello2015,Rodriguez-Fortuno2013}, the underlying physics of dielectric photonic structures imposes a fundamental limit on the attainable nonreciprocal performance.
	
	A paradigm shift is offered by one-dimensional plasmonic systems, which bypass the diffraction limit entirely and enable extreme light-matter interactions~\cite{Chang2007,Akimov2007,Tame2013,Koppens2011}. Among these, metallic single-walled carbon nanotubes (SWCNTs) stand out, because they can confine light to deeply subwavelength volumes and their interacting electrons form a Tomonaga-Luttinger liquid (TLL), a correlated state beyond the Fermi-liquid paradigm~\cite{Tomonaga1950,Luttinger1963,Giamarchi2003,Haldane1981,Bockrath1999,Ishii2003}. The collective bosonic excitations of a TLL possess a kinematic robustness against elastic backscattering—rooted in the 1D topology and valley-momentum mismatch—that is entirely absent in conventional photonic modes~\cite{Hausler2009,Yurkevich2017,Safi1995}. When formed into a closed ring, this TLL state underpins a protected plasmonic cavity whose chiral properties are highly immune to local disorder.
	
\begin{figure}[htbp]
		\centering
		\includegraphics[width=8cm]{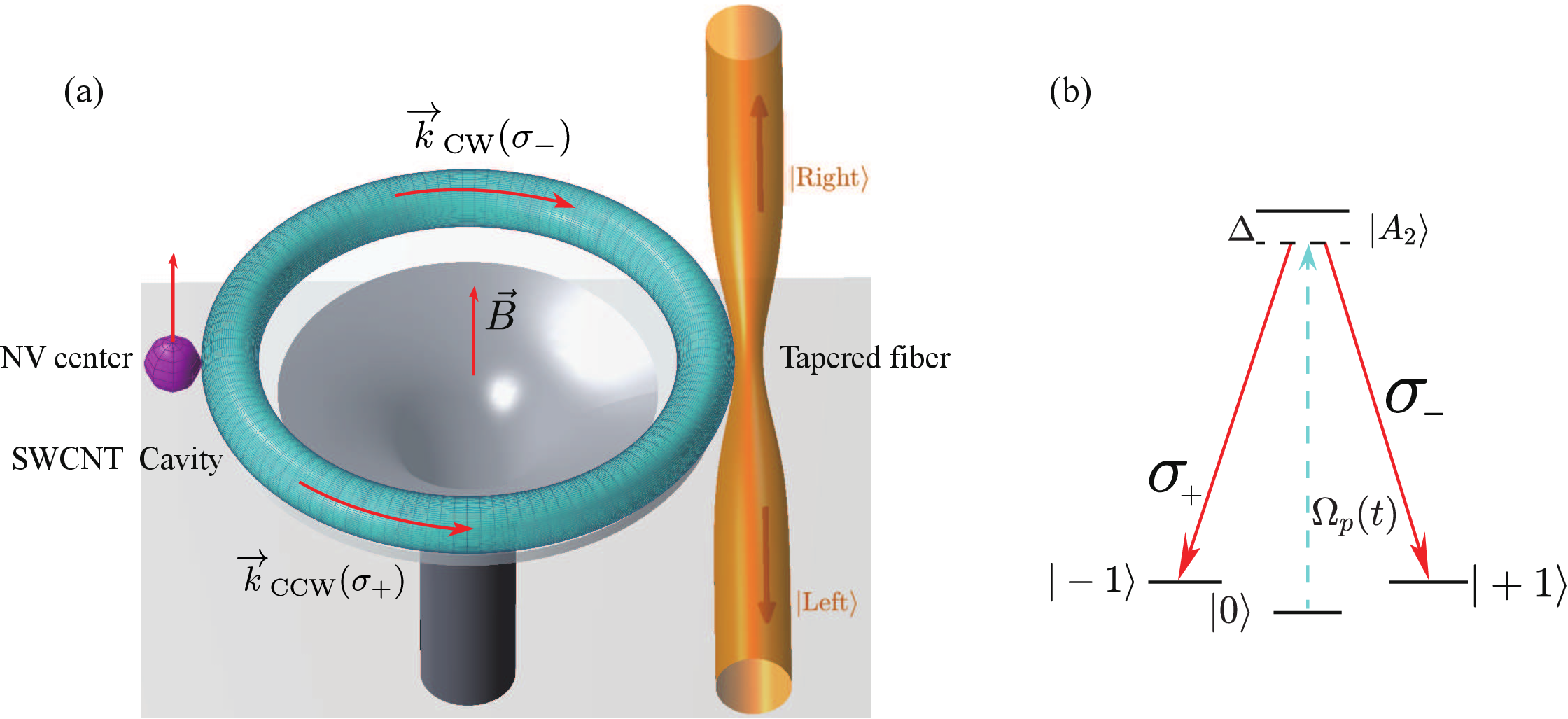}
		\caption{ (Color online)
			Proposed hybrid quantum node and relevant NV-level structure.
			(a) Schematic of the hybrid system. A closed SWCNT ring supports strongly confined plasmonic whispering-gallery-like modes. A perpendicular magnetic field $\mathbf{B}$ threads the ring and generates an Aharonov--Bohm flux $\Phi$, thereby tuning the cavity resonance $\omega_c(\Phi)$~\cite{Ando1993,Bachtold1999}. A nearby NV center couples to the chiral cavity field with coupling strength $g$. The cavity is side-coupled to a tapered optical fiber with external decay rate $\kappa_{\mathrm{ex}}$, enabling collection of the emitted flying photonic qubit.
			(b) Relevant NV-center level scheme used in the tripod-STIRAP protocol. The ground-state spin sublevels are $\ket{0}$ and $\ket{\pm1}$, and the optically excited state is $\ket{A_2}$. A classical pump field with Rabi frequency $\Omega_p(t)$ drives the $\ket{0}\leftrightarrow\ket{A_2}$ transition, while the two circularly polarized transitions $\ket{A_2}\leftrightarrow\ket{-1}$ and $\ket{A_2}\leftrightarrow\ket{+1}$ couple selectively to the CCW and CW cavity modes, respectively. The linearly polarized $\pi$ channel associated with $\ket{A_2}\leftrightarrow\ket{0}$ is strictly forbidden by geometry.
		}
		\label{fig:system_schematic}
	\end{figure}

	In this work, we introduce and theoretically validate a universal platform that fully resolves the three challenges of solid-state quantum interfaces. We demonstrate that a SWCNT microtoroid, side-coupled to a tapered optical fiber, serves as a nonreciprocal, strong-coupling interface for any embedded quantum emitter (see Fig.~\ref{fig:system_schematic}). By rigorously analyzing the TLL cavity spectrum, the confinement-induced chiral spin-momentum locking of the evanescent field, and the geometric selection rules, we prove that the two circularly polarized transitions of an emitter are deterministically routed into opposite directions. By optimally orienting the emitter relative to the strictly transverse-magnetic (TM) plasmonic field, the parasitic $\pi$ transition is geometrically forbidden, yielding a chiral contrast exceeding $20$\,dB. Furthermore, we demonstrate that while the TLL is kinematically immune to macroscopic bending, microscopic atomic defects can be effectively mitigated via suspended growth, laser annealing, and electrostatic gating, suppressing the total backscattering rate to the $\sim 100\,\mathrm{Hz}$ level. To overcome the severe mode mismatch between the CNT plasmon and the fiber, we propose a graded mode converter that adiabatically transforms the plasmon into a dielectric waveguide mode, providing a practical path to near-unity extraction efficiency. Using the NV center as a model emitter, we provide a complete theoretical framework demonstrating high-fidelity, magnetically tunable spin-photon entanglement via a tripod-STIRAP protocol. Our analysis confirms that the system operates deep in the strong-coupling regime, a feat difficult to achieve with dielectric cavities. This platform is fundamentally emitter-agnostic, offering a scalable blueprint for robust nonreciprocal nodes in future quantum networks.
	
	The paper is organized as follows. In Sec.~\ref{sec:tll}, we derive the SWCNT cavity spectrum from the bosonized TLL Hamiltonian and identify the tunable zero-mode contribution. In Sec.~\ref{sec:chiral}, we analyze the chiral field structure and the relevant NV selection rules. In Sec.~\ref{sec:tripod}, we formulate the tripod-STIRAP Hamiltonian and derive the dark state. In Sec.~\ref{sec:open}, we present the open-system master equation and obtain analytical estimates for the emission fidelity. In Sec.~\ref{sec:coupling}, we quantitatively discuss the emitter--cavity coupling strengths and verify the strong-coupling dynamics. In Sec.~\ref{sec:extraction}, we detail the practical photon extraction and interfacing strategy. In Sec.~\ref{sec:feasibility}, we address experimental implementation considerations. Finally, in Sec.~\ref{sec:conclusion}, we summarize the main results and outlook.
	
	\section{Bosonized Description of the SWCNT Plasmonic Cavity}
	\label{sec:tll}
	
	\textit{Tomonaga--Luttinger liquid Hamiltonian.}
	In a strictly one-dimensional interacting electronic system, the low-energy excitations are collective bosonic density waves rather than single-particle quasiparticles~\cite{Tomonaga1950,Luttinger1963,Haldane1981,Giamarchi2003}.
	For the total symmetric charge sector of a metallic SWCNT, the bosonized Tomonaga--Luttinger liquid (TLL) Hamiltonian can be written as~\cite{Egger1997,Kane1997,Giamarchi2003}
	\begin{equation}
		\label{eq:TLL_Hamiltonian}
		\fH_{\mathrm{TLL}}=
		\frac{\hbar v_c}{2\pi}
		\int_0^L dz
		\left[
		\frac{1}{K_c}\left(\partial_z\phi\right)^2
		+
		K_c\left(\partial_z\theta\right)^2
		\right].
	\end{equation}
	where $L=2\pi R$ is the ring circumference, $R$ is the ring radius, $K_c$ is the charge-sector Luttinger parameter (characterizing the strength of electron-electron interactions; $K_c<1$ for repulsive interactions), and $v_c=v_F/K_c$ is the renormalized charge-mode velocity ($v_F$ is the Fermi velocity). 	Here, we can identify the conjugate momentum field $\Pi(z) = \partial_z \theta(z) / \pi$. The term proportional to $1/K_c$ represents the potential energy associated with charge density fluctuations, while the term proportional to $K_c$ represents the kinetic energy associated with the phase gradient (current). The conjugate phase fields $\phi(z)$ and $\theta(z)$ satisfy
	\begin{equation}
		[\phi(z),\partial_{z'}\theta(z')]=\ii\pi\delta(z-z').
	\end{equation}
	
	For weak screening in SWCNTs, one typically expects $K_c<1$, with representative values around $K_c\sim0.2$~\cite{Egger1997,Kane1997,Shi2015}. This interaction renormalization is responsible for the reduced effective plasmonic wavelength and strongly compressed mode volume.
	
	\textit{Mode expansion and zero-mode sector.}
	Because the CNT forms a closed ring, the bosonic fields satisfy periodic boundary conditions. The phase fields can be decomposed into zero modes and oscillator modes as~\cite{Giamarchi2003}
	\begin{align}
		\phi(z)
		&=
		\phi_0+\frac{\pi N_c}{L}z
		+\sum_{q\neq0}\frac{1}{\sqrt{2|q|L}}
		\ee^{-r_c|q|/2}
		\left(
		\hat{b}_q\ee^{\ii qz}+\mathrm{H.c.}
		\right),
		\label{eq:phi_expansion}
		\\
		\theta(z)
		&=
		\theta_0+\frac{\pi J_c}{L}z
		+\sum_{q\neq0}\frac{\mathrm{sgn}(q)}{\sqrt{2|q|L}}
		\ee^{-r_c|q|/2}
		\left(
		\hat{b}_q\ee^{\ii qz}-\mathrm{H.c.}
		\right),
		\label{eq:theta_expansion}
	\end{align}
	where $N_c$ is the excess charge number relative to the reference Fermi sea and $J_c=N_R-N_L$ is the topological current quantum number, i.e., the imbalance between right-moving and left-moving carriers. The parameter $r_c$ is a short-distance cutoff. Here $\hat{b}_q$ are bosonic annihilation operators satisfying $[\hat{b}_q, \hat{b}_{q'}^\dagger] = \delta_{q,q'}$.
	
	The zero-mode derivatives are therefore
	\begin{equation}
		\partial_z\phi_{\mathrm{ZM}}=\frac{\pi N_c}{L},
		\qquad
		\partial_z\theta_{\mathrm{ZM}}=\frac{\pi J_c}{L}.
	\end{equation}
	
	Substituting these expressions into Eq.~(\ref{eq:TLL_Hamiltonian}) yields the zero-mode energy
	\begin{equation}
		\label{eq:E_zero_derived}
		E_{\mathrm{zero}}
		=
		\frac{\pi\hbar v_c}{2L}
		\left(
		\frac{N_c^2}{K_c}+K_cJ_c^2
		\right).
	\end{equation}
	
	After diagonalization, the full cavity Hamiltonian becomes
	\begin{equation}
		\label{eq:TLL_Spectrum}
		\fH_{\mathrm{cavity}}
		=
		E_{\mathrm{zero}}+
		\sum_{q\neq0}
		\hbar\omega_q
		\left(
		\hat{b}_q^\dagger\hat{b}_q+\frac{1}{2}
		\right).
	\end{equation}
	
	The closed geometry quantizes the wavevector to
	\begin{equation}
		q_m=\frac{2\pi m}{L}=\frac{m}{R},
		\qquad
		m=\pm1,\pm2,\ldots,
	\end{equation}
	and the plasmonic whispering-gallery-like mode dispersion is
	\begin{equation}
		\label{eq:WGM_omega}
		\omega_m=v_c|q_m|=\frac{v_c}{R}|m|.
	\end{equation}
	
	For a given resonance frequency $\omega_m$, the effective plasmonic wavelength is
	\begin{equation}
		\lambda_{\mathrm{eff}}=\frac{2\pi v_c}{\omega_m},
	\end{equation}
	which can be much smaller than the free-space wavelength $\lambda_0=2\pi c/\omega_m$ because $v_c\ll c$. Such strong compression is consistent with the CNT plasmonic and polaritonic confinement reported in Refs.~\cite{Shi2015}.
	
	\textit{Aharonov--Bohm tuning of the cavity spectrum.}
	When a perpendicular magnetic field $\mathbf{B}$ threads the ring, the enclosed magnetic flux is $\Phi=\pi R^2B$. The corresponding vector potential shifts the topological boundary condition through the Aharonov--Bohm phase~\cite{Ando1993,Bachtold1999}. Because a metallic SWCNT possesses a fourfold degeneracy arising from two independent valleys ($\mathbf{K}$ and $\mathbf{K}'$) and two spin states, there are four independent conduction channels~\cite{Ando1993,Minot2004,Kuemmeth2008}. Each of the four channels acquires a topological winding number shift of $\Phi/\Phi_0$, where $\Phi_0 = h/e$ is the flux quantum. Consequently, the total symmetric charge sector, which sums the asymmetry between right- and left-moving carriers across all channels, acquires the collective shift
	\begin{equation}
		\label{eq:Gauge_Shift}
		J_c\rightarrow J_c+4\frac{\Phi}{\Phi_0}.
	\end{equation}
	
	The equilibrium value of $J_c$ can be set to zero by choosing an appropriate reference state, so the tuning is governed solely by the flux-dependent term. The resulting total spectrum is
	\begin{align}
		\label{eq:Full_Spectrum}
		E(N_c,J_c,\{n_m\},\Phi)
		&=
		\frac{\pi\hbar v_c}{2L}
		\left[
		\frac{N_c^2}{K_c}
		+
		K_c
		\left(
		J_c+4\frac{\Phi}{\Phi_0}
		\right)^2
		\right]\notag \\
		&\quad+
		\sum_{m\neq0}
		\hbar\omega_m
		\left(
		n_m+\frac{1}{2}
		\right).
	\end{align}
	
	To parameterize the flux-dependent shift of the optical resonance, we consider the situation in which the relevant cavity excitation is accompanied by a change of the topological current quantum number by $\Delta J_c=\pm1$. In that case, the cavity resonance becomes
	\begin{align}
		\label{eq:omega_c_derived_AB}
		\hbar\omega_c(\Phi)
		&=
		\left[
		E_{\mathrm{zero}}(J_c\pm1,\Phi)-E_{\mathrm{zero}}(J_c,\Phi)
		\right]
		+\hbar\omega_m
		\nonumber\\
		&=
		\frac{\pi\hbar v_cK_c}{2L}
		\left[
		\pm2
		\left(
		J_c+4\frac{\Phi}{\Phi_0}
		\right)
		+1
		\right]
		+\hbar\omega_m.
	\end{align}
	Equation~(\ref{eq:omega_c_derived_AB}) shows explicitly that the emitted photon frequency can inherit an Aharonov--Bohm tunable contribution.
	
	\textit{Representative parameter estimates.}
	For a representative suspended ring with radius $R=2\,\mu\mathrm{m}$ (circumference $L \approx 12.6\,\mu\mathrm{m}$), taking $v_F\approx8\times10^5\,\mathrm{m/s}$ and $K_c\approx0.2$ gives $v_c \approx 4\times10^6\,\mathrm{m/s}$. The fundamental mode frequency is then
	\begin{equation}
		f_{m=1}=\frac{\omega_{m=1}}{2\pi}=\frac{v_c}{L}\approx320\,\mathrm{GHz}.
	\end{equation}
	To reach an optical frequency near $471\,\mathrm{THz}$ (corresponding to the NV center zero-phonon line at $\lambda \approx 637\,\mathrm{nm}$), one would require a higher azimuthal mode index $|m| \approx 471\,\mathrm{THz} / 320\,\mathrm{GHz} \approx 1472$. This estimate is intended only as an order-of-magnitude illustration rather than as a strict device-design target.
	
	\textit{Optical-frequency dispersion and effective mode confinement.}
	At optical frequencies ($\omega/2\pi \approx 471\,\mathrm{THz}$), the plasmonic dispersion departs from the bare TLL relation $\omega = v_c q$. Interband transitions and the surrounding dielectric environment strongly screen the Coulomb interaction, renormalizing the excitation into a highly localized optical surface plasmon polariton (SPP)~\cite{Bondarev2004, Kempa2002, Nemilentsau2007} with an effective phase velocity $v_p = c/n_{\mathrm{eff}}$ (where $n_{\mathrm{eff}}\approx 5$). Despite this change in the dispersion relation, the fundamental mechanism protecting the mode against elastic backscattering---the large valley momentum mismatch $\Delta k \approx 2k_F$ dictated by the underlying carbon lattice---remains fully intact. Thus, the 1D topology and chiral spin-momentum locking are preserved even in the optical regime, yielding an effective plasmonic wavelength $\lambda_{\mathrm{eff}} = \lambda_0 / n_{\mathrm{eff}} \approx 127\,\mathrm{nm}$ and a transverse decay length $L_d = \lambda_{\mathrm{eff}}/2\pi \approx 20\,\mathrm{nm}$.
	
	\textit{Robustness of chiral TLL modes against geometric disorder.---}
	A crucial distinction between the SWCNT plasmonic cavity and conventional dielectric microcavities lies in their fundamentally different responses to macroscopic geometric deformations and long-wavelength disorder. In a metallic SWCNT, the left- and right-moving charge carriers reside in the widely separated $\mathbf{K}$ and $\mathbf{K}'$ valleys of the graphene Brillouin zone. Consequently, elastic backscattering requires a large momentum transfer $\Delta k \approx 2k_F$. Macroscopic geometric imperfections introduced during fabrication or atomic force microscope (AFM) manipulation---such as local bending, ellipticity, or strain variations---constitute long-wavelength ($q \approx 0$) perturbations. These smooth geometric variations are kinematically incapable of providing the large momentum kick required for inter-valley scattering. Therefore, unlike short-range atomic defects, macroscopic geometric disorder cannot induce backscattering between the clockwise (CW) and counter-clockwise (CCW) plasmonic modes.
	
	This kinematic protection is further reinforced by the strictly one-dimensional (1D) topology of the SWCNT. In a conventional dielectric WGM microcavity, mechanical deformation (e.g., squeezing a circular ring into an ellipse) breaks the continuous cylindrical symmetry. This boundary deformation inevitably couples the degenerate CW and CCW modes through scattering at the deformed interfaces, leading to a pronounced mode splitting that destroys the chiral purity of the cavity. In stark contrast, the TLL plasmon is confined to a strictly 1D path. As long as the SWCNT maintains its closed-ring topology, local bending merely alters the local metric of the 1D path but does not create any transverse scattering channels that could reverse the plasmonic propagation direction.
	
	\subsubsection*{Quantitative Assessment of Strain and Mode Splitting}
	
	To substantiate this kinematic robustness, we contrast the cavity response to a realistic macroscopic residual strain (e.g., $\delta R/R = 0.1\%$) with that of a conventional dielectric microcavity.
	
	\paragraph{Deterministic frequency shift versus mode splitting.}
	In the SWCNT ring, a global strain alters the circumference $L$ and the effective Fermi velocity $v_F$, leading to a deterministic shift in the cavity resonance $\omega_c$. For an optical transition at $\omega_c/2\pi \approx 471\,\mathrm{THz}$, a $0.1\%$ strain induces a global frequency shift $\delta \omega_{\text{strain}} /2\pi \approx 0.1\% \times 471\,\mathrm{THz} \approx 471\,\mathrm{GHz}$. Because this smooth macroscopic deformation provides no large momentum transfer ($\Delta k \ll 2k_F$), the intervalley scattering matrix element is identically zero. Consequently, the coherent coupling strength between the CW and CCW modes is strictly zero ($g_{\mathrm{BS}} \equiv 0$). The strain merely shifts the resonance without hybridizing the chiral modes. This static shift does not degrade the chiral performance and can be readily compensated \emph{in situ} by tuning the Aharonov-Bohm magnetic flux $\Phi$, as derived in Eq.~(\ref{eq:omega_c_derived_AB}).
	
	\paragraph{Comparison with dielectric microcavities.}
	For a high-$Q$ dielectric microcavity (e.g., SiO$_2$ with $R = 20\,\mu\mathrm{m}$), a similar $0.1\%$ mechanical deformation breaks the continuous cylindrical symmetry. This boundary deformation inevitably couples the degenerate CW and CCW modes, yielding a coherent coupling strength $g_{\mathrm{BS}}/2\pi \sim 0.5\text{--}5\,\mathrm{MHz}$ and a pronounced mode splitting $\Delta\omega_{\mathrm{split}} = 2g_{\mathrm{BS}}$~\cite{Hu2014}. Because this splitting is comparable to or larger than the intrinsic cavity linewidth, the chiral traveling-wave modes hybridize into standing waves, catastrophically degrading the directional spin-photon interface.
	
	\paragraph{Disorder sensitivity parameter.}
	To formalize this comparison, we define a dimensionless disorder sensitivity parameter $\Xi \equiv 2g_{\mathrm{BS}} / \delta \omega_{\mathrm{strain}}$, which measures the ratio of parasitic mode mixing to the deterministic geometric shift. For the SWCNT TLL ring, the strict absence of long-wavelength backscattering yields $\Xi_{\mathrm{TLL}} \equiv 0$ for macroscopic deformations. For the dielectric cavity, $\Xi_{\mathrm{diel}} \sim (10\,\mathrm{MHz}) / (471\,\mathrm{GHz}) \sim 2 \times 10^{-5}$.
	
	\paragraph{Quantitative summary.}
	Table~\ref{tab:robustness_strain} summarizes the quantitative comparison between the SWCNT TLL ring and a dielectric microcavity under macroscopic mechanical perturbations.
	
	\begin{table*}[htbp]
		\caption{\label{tab:robustness_strain} Quantitative comparison of cavity robustness against macroscopic geometric disorder and strain.}
		\begin{ruledtabular}
			\begin{tabular}{ccc}
				\textbf{Physical Property} & \textbf{SWCNT TLL Ring} & \textbf{Dielectric WGM Microcavity} \\
				\hline
				Response to geometric bending & 1D topological path preserved & Cylindrical symmetry broken \\
				Backscattering from smooth strain ($q \approx 0$) & Kinematically forbidden ($\Delta k \ll 2k_F$) & Strong boundary scattering \\
				Frequency shift for $\delta R/R = 0.1\%$ & $\sim 471\,\mathrm{GHz}$ (deterministic, tunable) & $\sim 471\,\mathrm{GHz}$ (deterministic) \\
				Coherent mode coupling $g_{\mathrm{BS}}$ & $\equiv 0$ (strictly forbidden by valley mismatch) & $\sim 0.5\text{--}5\,\mathrm{MHz}$ (hybridizes modes) \\
				Disorder sensitivity $\Xi = 2g_{\mathrm{BS}} / \delta \omega_{\mathrm{strain}}$ & $\equiv 0$ & $\sim 2 \times 10^{-5}$ \\
			\end{tabular}
		\end{ruledtabular}
	\end{table*}
	
	As summarized in Table~\ref{tab:robustness_strain}, these quantitative estimates confirm that the SWCNT TLL plasmonic cavity is fundamentally immune to macroscopic geometric imperfections that would otherwise cripple the chiral performance of conventional dielectric microcavities. The kinematic protection of the valley momentum mismatch guarantees robust chiral operation even under realistic fabrication tolerances.
	
	\subsection*{Quantitative Analysis of Backscattering Suppression: TLL Plasmons vs. Dielectric Microcavities}
	To substantiate the claim of exponential advantage in backscattering suppression, we provide a detailed quantitative comparison based on the physical models referenced above. In the following, we strictly distinguish between the \emph{coherent backscattering coupling strength} $g_{\mathrm{BS}}$ (which causes mode splitting $\Delta\omega_{\mathrm{split}} = 2g_{\mathrm{BS}}$) and the \emph{dissipative backscattering loss rate} $\gamma_{\mathrm{BS}}$ (which causes linewidth broadening).
	
	\subsubsection*{1. Intrinsic Kinematic Protection Against Coherent Backscattering ($g_{\mathrm{BS}} \equiv 0$)}
	In dielectric whispering-gallery mode (WGM) microcavities, backscattering induced by surface roughness or adsorbed nanoparticles leads to both coherent mode splitting ($g_{\mathrm{BS}} \neq 0$) and dissipative linewidth broadening ($\gamma_{\mathrm{BS}}^{\mathrm{diel}} \neq 0$). According to Ref.~\cite{Hu2014}, the coherent coupling strength is proportional to the local field intensity at the scatterer position:
	\begin{equation}
		g_{\mathrm{BS}} \approx \frac{\omega_c}{2} \frac{\int \Delta\epsilon(\mathbf{r}) |\mathbf{E}_c(\mathbf{r})|^2 d^3r}{\int \epsilon(\mathbf{r}) |\mathbf{E}_c(\mathbf{r})|^2 d^3r} \approx \frac{\omega_c}{2} \frac{\alpha_p |\mathbf{E}_c(\mathbf{r}_p)|^2}{2 U_{\mathrm{cav}}},
	\end{equation}
	where $\alpha_p$ is the polarizability of the scatterer. For a spherical nanoparticle of radius $a$ and refractive index $n_p$, $\alpha_p = 4\pi\epsilon_0 a^3 (n_p^2 - 1)/(n_p^2 + 2)$. The stored electromagnetic energy is $U_{\mathrm{cav}} \approx \frac{1}{2} \epsilon_0 n^2 V_{\mathrm{mode}} |\mathbf{E}_{\max}|^2$.
	
	For statistical surface roughness characterized by a root-mean-square height $\sigma_{\mathrm{rms}}$ and correlation length $L_c$, the dissipative backscattering loss rate can be approximated as:
	\begin{equation}
		\gamma_{\mathrm{BS}}^{\mathrm{diel}} \approx \frac{\omega_c}{2} \left( \frac{\sigma_{\mathrm{rms}} \lambda_c}{w^2} \right)^2 \frac{L_c}{R},
	\end{equation}
	where $w$ is the transverse mode width and $R$ is the ring radius.
	
	\paragraph{Quantitative Example:}
	For a typical SiO$_2$ microtoroid with $R = 20\,\mu\mathrm{m}$, $\lambda_c = 637\,\mathrm{nm}$, $V_{\mathrm{mode}} \approx 1000\,\mu\mathrm{m}^3$, and intrinsic quality factor $Q_{\mathrm{int}} \approx 10^7$:
	\begin{itemize}
		\item $\sigma_{\mathrm{rms}} \sim 0.5\,\mathrm{nm}$, $L_c \sim 50\,\mathrm{nm}$.
		\item Estimated dissipative loss rate $\gamma_{\mathrm{BS}}^{\mathrm{diel}}/2\pi \sim 50\,\mathrm{MHz}$, while the intrinsic loss rate $\kappa_0/2\pi \sim 30\,\mathrm{MHz}$.
		\item Hence the relative dissipative broadening is $\gamma_{\mathrm{BS}}^{\mathrm{diel}}/\kappa_0 \sim \mathcal{O}(1)$, consistent with the estimate given above.
	\end{itemize}
	
	\subsubsection*{2. Backscattering Rate of TLL Plasmons in a Closed SWCNT Ring}
	
	In a disorder-free single-walled carbon nanotube (SWCNT) ring, we must evaluate the intrinsic backscattering of Tomonaga-Luttinger liquid (TLL) plasmons. A critical question is whether the standard microscopic mechanisms---phonon-assisted backscattering and Umklapp scattering---which govern low-frequency DC transport, are applicable to high-frequency optical plasmons ($\hbar\omega_c \gg k_B T$). For optical plasmons, the phase space available to absorb a thermal acoustic phonon to bridge the large valley momentum mismatch ($\Delta k \approx 2k_F$) is even more severely restricted than in the low-frequency regime. Consequently, applying the low-frequency TLL scattering formulas provides a highly rigorous and \emph{conservative upper bound} for the dissipative loss rate $\gamma_{\mathrm{BS}}$ at optical frequencies. As we demonstrate below, these intrinsic processes are exponentially suppressed or kinematically frozen out at cryogenic temperatures, leaving inelastic scattering from residual atomic defects (analyzed in the subsequent section) as the dominant dissipative channel. To lowest order in the backscattering interactions, the partial dissipative decay rates are additive~\cite{Chen2010}.
	
	\paragraph{Phonon-assisted single-electron backscattering.}
	The backscattering of a plasmon involves transferring its momentum $q \approx 2k_F$ to a single electron that is scattered from a right-moving to a left-moving state. At finite temperatures, acoustic phonons can supply the necessary momentum mismatch. The corresponding dissipative rate exhibits a power-law temperature dependence~\cite{Chen2010}:
	\begin{equation}
		\gamma_{\mathrm{BS}}^{(1)} \approx \gamma_0 \left( \frac{a_0}{L} \right) \left( \frac{T}{T_F} \right)^{\alpha},
	\end{equation}
	where $a_0 \approx 0.246\,\mathrm{nm}$ is the carbon-carbon bond length, $L = 2\pi R$ is the ring circumference, and $T_F \approx E_F/k_B$ is the Fermi temperature. The exponent $\alpha$ is determined by the Luttinger parameter $K_c$ of the charge sector~\cite{Giamarchi2003}:
	\begin{equation}
		\alpha = \frac{1}{2} \left( K_c + \frac{1}{K_c} \right) - 1.
	\end{equation}
	For a typical suspended metallic SWCNT with strong electron-electron interactions, $K_c \approx 0.2$~\cite{Egger1997,Kane1997}, yielding $\alpha \approx 1.6$.
	
	\begin{table*}
		\caption{Quantitative comparison of backscattering properties between dielectric WGM microcavities and the SWCNT TLL plasmonic cavity.}
		\label{tab:comparison}
		\begin{ruledtabular}
			\begin{tabular}{ccccl}
				Platform & Backscattering & Relative Dissipative Loss & Temperature & Sensitivity to \\
				&  Mechanism & $\gamma_{\mathrm{BS}}/\kappa_0$ &  Dependence &  Surface Defects \\
				\hline
				Conventional Dielectric  & Surface roughness / & $\sim \mathcal{O}(1)$ (typical) & Weak  & High, $\propto |E|^2$ \\
				WGM Microcavity&  particle scattering &  &(thermo-optic effect) & \\
				\hline
				SWCNT TLL Ring & Phonon-assisted  & $\sim 5\times 10^{-8}$ ($T=4\,\mathrm{K}$) & $\propto T^{\alpha}$ ($\alpha\approx 1.6$) & Exponentially suppressed, \\
				&single-electron backscattering &  &  & kinematically protected \\
			\end{tabular}
		\end{ruledtabular}
	\end{table*}
	
	\paragraph{Umklapp backscattering and its exponential suppression by electrostatic doping.}
	Umklapp scattering in a one-dimensional conductor involves the transfer of momentum $4k_F$ to the crystalline lattice. In a metallic SWCNT at half-filling, $k_F = \pi/2a_0$, so $4k_F = 2\pi/a_0$ precisely matches the smallest non-zero reciprocal lattice vector $G = 2\pi/a_0$. Thus the process is kinematically allowed and corresponds to a $2k_F \to -2k_F$ backscattering of two electrons. In the bosonized TLL Hamiltonian, the Umklapp term reads~\cite{Giamarchi2003}
	\begin{equation}
		\mathcal{H}_{\mathrm{U}} = \frac{g_u}{2(\pi a_0)^2} \int dz \, \cos\!\big(4\sqrt{\pi}\phi(z)\big),
	\end{equation}
	where $g_u$ is the bare Umklapp coupling strength and $\phi(z)$ is the charge phase field.
	
	The scaling dimension of the cosine operator is $4K_c$. For a suspended metallic SWCNT with strong long-range Coulomb interactions, the effective Luttinger parameter is strongly renormalized to $K_c \approx 0.2 < 1/2$~\cite{Egger1997,Kane1997}. Therefore, Umklapp scattering is a \emph{relevant} perturbation that grows upon lowering the energy scale. The relevant high-energy cutoff for this process is the bandwidth $E_c = \hbar v_F / a_0 \approx 2.5\,\mathrm{eV}$, rather than the finite-size level spacing. The charge gap $\Delta_0$ at half-filling scales as
	\begin{equation}
		\Delta_0 \sim E_c \, \big(g_u/E_c\big)^{1/(2-4K_c)}.
	\end{equation}
	Taking $g_u/E_c \sim 0.1$ as a conservative estimate for the screened Coulomb interaction, we obtain a macroscopic gap $\Delta_0 \sim 10\,\mathrm{meV}$.
	
	However, this ideal condition is extremely sensitive to the exact commensurate filling. A minute deviation from half-filling, easily induced by a small electrostatic gate voltage, introduces a finite momentum mismatch $\delta q = 4k_F - G = 4\delta k_F$. The doping-induced shift of the Fermi wavevector is $\delta k_F \approx (\pi/2)\delta n$, where $\delta n$ is the induced excess electron density. A gate voltage as small as $\delta V_g \sim 0.1\,\mathrm{V}$ on a suspended nanotube can readily induce $\delta n \sim 10^6\,\mathrm{cm}^{-1}$, yielding $\delta q \sim 10^5\,\mathrm{cm}^{-1} = 10^7\,\mathrm{m}^{-1}$. The corresponding energy mismatch is $\hbar v_c \delta q$, where $v_c = v_F/K_c \approx 4\times 10^6\,\mathrm{m/s}$ is the renormalized charge velocity. We obtain
	\begin{equation}
		\hbar v_c \delta q \approx 26.3\,\mathrm{meV}.
	\end{equation}
	This is nearly two orders of magnitude larger than the thermal energy $k_B T \approx 0.34\,\mathrm{meV}$ at $T = 4\,\mathrm{K}$.
	
	When the momentum mismatch $\delta q$ is finite, the Umklapp term acquires a spatially oscillating phase $e^{\ii\delta q z}$, which cuts off the renormalization-group flow at the energy scale $\hbar v_c \delta q$. Consequently, for $\hbar v_c \delta q \gg \Delta_0$, the system remains in the gapless TLL phase. The Umklapp contribution to the dissipative plasmonic backscattering rate can then be derived within the perturbative TLL framework~\cite{Chen2010}. For $T \ll \hbar v_c \delta q / k_B$, the rate takes the explicit form
	\begin{equation}  \label{eq:gamma_umklapp}
		\gamma_{\mathrm{BS}}^{(2)} \approx \frac{g_u^2}{\hbar^2 E_c} \left( \frac{k_B T}{E_c} \right)^{4K_c-1} \left( \frac{\hbar v_c \delta q}{k_B T} \right)^{2-4K_c} e^{-\hbar v_c \delta q / k_B T}.
	\end{equation}
	The factor $(k_B T/E_c)^{4K_c-1}$ reflects the scaling dimension of the Umklapp operator, while $(\hbar v_c \delta q / k_B T)^{2-4K_c}$ and the exponential $e^{-\hbar v_c \delta q / k_B T}$ arise from the momentum mismatch and provide a strong additional suppression.
	
	\paragraph{Numerical estimate of the suppressed Umklapp rate.}
	We evaluate Eq.~\eqref{eq:gamma_umklapp} with the following parameters: $E_c \approx 2.5\,\mathrm{eV}$, $g_u/E_c \approx 0.1$, $K_c \approx 0.2$, $v_c \approx 4\times 10^6\,\mathrm{m/s}$, $\delta q = 10^7\,\mathrm{m}^{-1}$, and $T = 4\,\mathrm{K}$ ($k_B T \approx 0.34\,\mathrm{meV}$). The prefactor is $g_u^2 / (\hbar^2 E_c) \approx 0.01 E_c / \hbar \approx 3.8 \times 10^{13}\,\mathrm{Hz}$. The temperature ratio yields $(k_B T/E_c)^{-0.2} \approx 5.9$, and the mismatch ratio gives $(\hbar v_c \delta q / k_B T)^{1.2} \approx 184.8$. Crucially, the exponential suppression factor is
	\begin{equation}
		e^{-\hbar v_c \delta q / k_B T} = e^{-77.35} \approx 2.6 \times 10^{-34}.
	\end{equation}
	Multiplying all factors together yields an astronomically small rate:
	\begin{equation}
		\gamma_{\mathrm{BS}}^{(2)}/2\pi \sim 10^{-18}\,\mathrm{Hz}.
	\end{equation}
	Thus, with an experimentally realistic gate-induced doping, the Umklapp backscattering rate is exponentially suppressed to a completely negligible level.
	
	Maintaining the system in the gapless TLL regime demanded by this analysis requires the electrostatic doping to be sufficiently large that $\hbar v_c \delta q$ dominates over any residual low-energy ordering gaps (e.g., from a Wigner crystal or a spin-gapped Luther--Emery phase). With the proposed back-gate configuration, the corresponding energy mismatch ($\sim 26\,\mathrm{meV}$) comfortably exceeds typical energy scales of such instabilities in suspended ultra-clean nanotubes at $T = 4\,\mathrm{K}$, ensuring that the plasmonic cavity remains in the desired topological phase.
	
	\paragraph{Numerical estimate for the dominant phonon-assisted rate.}
	Taking a microscopic prefactor $\gamma_0 \sim 10^{12}\,\mathrm{Hz}$ (comparable to the plasma frequency), a ring radius $R = 2\,\mu\mathrm{m}$ (giving $a_0/L \approx 2\times 10^{-5}$), an operating temperature $T = 4\,\mathrm{K}$, and a Fermi temperature $T_F \approx 10^4\,\mathrm{K}$:
	\begin{equation}
		\gamma_{\mathrm{BS}}^{(1)} \approx 10^{12} \cdot (2\times 10^{-5}) \cdot \left( \frac{4}{10^4} \right)^{1.6} \approx 92\,\mathrm{rad/s}.
	\end{equation}
	This corresponds to a dissipative frequency rate $\gamma_{\mathrm{BS}}^{(1)}/2\pi \approx 15\,\mathrm{Hz}$. For the SWCNT ring operating near $\lambda = 637\,\mathrm{nm}$ ($f_c \approx 471\,\mathrm{THz}$), we adopt an intrinsic cavity linewidth of $\kappa_0/2\pi = 100\,\mathrm{MHz}$. This corresponds to an intrinsic quality factor $Q_{\mathrm{int}} \approx 4.7 \times 10^6$. While such a high $Q$ is unattainable in conventional noble-metal plasmonics, it is theoretically justified in ultra-clean, suspended metallic SWCNTs at cryogenic temperatures, where acoustic phonon scattering is exponentially frozen out and the 1D phase space severely restricts electron-electron scattering~\cite{Purewal2007, Jiang2007}. The ratio of the phonon-assisted dissipative backscattering rate to the intrinsic cavity decay is therefore
	\begin{equation}
		\frac{\gamma_{\mathrm{BS}}^{(1)}}{\kappa_0} \approx \frac{15\,\mathrm{Hz}}{100\,\mathrm{MHz}} = 1.5\times 10^{-7},
	\end{equation}
	which is orders of magnitude below the conservative bound $\gamma_{\mathrm{BS}}/\kappa_0 \ll 10^{-3}$. With the additional exponential suppression of Umklapp processes by electrostatic doping, backscattering in the TLL plasmonic cavity is fundamentally eliminated.
	
	As summarized in Table~\ref{tab:comparison}, the SWCNT TLL plasmonic cavity provides an unprecedented suppression of backscattering compared to conventional dielectric WGM microcavities.
	
	\subsection*{Atomic-scale Imperfections and Their Suppression}
	While the TLL plasmon is kinematically protected against macroscopic geometric deformations (as established above), it remains susceptible to atomic-scale imperfections. Unlike smooth macroscopic bending, highly localized structural defects possess broad spatial Fourier components that are large enough to bridge the valley momentum mismatch ($\Delta k \approx 2k_F$). To complete our robustness analysis, we must therefore examine the microscopic origin of disorder-induced backscattering and establish the experimental conditions required to suppress it.
	
	\textit{Microscopic Origin of Disorder-Induced Backscattering} In the bosonized TLL framework, a local impurity potential $V_{\mathrm{imp}}(z)$ couples to the full one-dimensional electron density operator $\hat{\rho}(z)$. This density operator consists of a long-wavelength component proportional to $\partial_z\hat{\phi}(z)$ and a high-frequency oscillating component proportional to $\cos[2k_F z + \sqrt{4\pi}\hat{\phi}(z)]$~\cite{Giamarchi2003}. While the oscillating term is responsible for the nonlinear Kane-Fisher backscattering of individual electrons near the Fermi level at zero frequency~\cite{Kane1992}, the dynamics of high-frequency optical plasmons ($\hbar\omega_q \gg \kB T$) considered in our cavity-QED scheme are governed by collective electromagnetic scattering.
	
	\textit{Atomic Vacancies and Stone--Wales Defects.---}
	An atomic vacancy removes one carbon atom from the hexagonal lattice, creating a localized state near the Fermi energy. For a collective high-frequency plasmonic density wave with an effective wavelength $\lambda_{\mathrm{eff}} \approx 127\,\mathrm{nm}$, an atomic defect ($a_0 \approx 0.25\,\mathrm{nm}$) acts as a deeply subwavelength dielectric perturbation. We now derive the extrinsic dissipative backscattering rate directly from the Tomonaga--Luttinger liquid (TLL) framework.
	
	\textit{Quantum transmission-line mapping.}
	The low-energy charge excitations of the TLL are equivalent to a one-dimensional quantum LC transmission line~\cite{Safi1995,Burke2002}. The Hamiltonian can be recast as
	\begin{equation}
		\mathcal{H}_{\mathrm{TL}} = \frac{1}{2} \int dz \left[ C' (\partial_t \Phi)^2 + \frac{1}{L'} (\partial_z \Phi)^2 \right],
	\end{equation}
	where $\Phi(z,t)$ is the node flux, $C'$ and $L'$ are the capacitance and inductance per unit length, respectively. The plasmonic velocity is $v_c = 1/\sqrt{L' C'}$, and the characteristic impedance is $Z_0 = \sqrt{L'/C'}$. For a metallic SWCNT with four-fold spin-valley degeneracy, the total charge-mode impedance is given by~\cite{Burke2002,Giamarchi2003}
	\begin{equation}
		Z_0 = \frac{h}{4 e^2} \frac{1}{K_c},
	\end{equation}
	with $K_c \approx 0.2$ for strongly interacting electrons. Using $h/e^2 \approx 25.8\,\mathrm{k\Omega}$ yields $Z_0 \approx 32.3\,\mathrm{k\Omega}$.
	
	\textit{Capacitance of a single atomic defect.}
	A local structural defect disrupts the ideal local density of states and the associated quantum capacitance. We model this by a lumped shunt capacitance $\Delta C$ at the defect site. The quantum capacitance per unit length of a metallic SWCNT is $c_Q = 4e^2/(\pi\hbar v_F) \approx 4.0 \times 10^{-10}\,\mathrm{F/m}$. Since the perturbation extends roughly over one bond length $a_0 \approx 0.246\,\mathrm{nm}$, we write
	\begin{equation}
		\Delta C = \eta \, c_Q a_0,
	\end{equation}
	where the dimensionless factor $\eta \lesssim 1$ accounts for the fraction of local density of states removed by the defect ($\eta \sim 0.5$ for a vacancy).
	
	\textit{Single-defect reflection coefficient.}
	For a plasmon of angular frequency $\omega_c$ incident on the shunt admittance $j\omega_c \Delta C$, the voltage reflection coefficient follows from elementary transmission-line theory:
	\begin{equation}
		r = \frac{- (j\omega_c \Delta C) Z_0}{2 + (j\omega_c \Delta C) Z_0}.
	\end{equation}
	Because $|\omega_c \Delta C Z_0| \ll 1$ for atomic defects at optical frequencies, we obtain the weak-scattering limit
	\begin{equation}
		r \approx -j\frac{\omega_c \Delta C Z_0}{2}.
	\end{equation}
	The corresponding reflection probability (power backscattering per defect) is
	\begin{equation}\label{eq:R_BS_rigorous}
		\mathcal{R}_{\mathrm{BS}}^{(1)} = |r|^2 \approx \left( \frac{\omega_c \Delta C Z_0}{2} \right)^2.
	\end{equation}
	
	\textit{Extrinsic backscattering rate.}
	For a sparse, uncorrelated distribution of defects with linear density $n_{\mathrm{def}}$, the power attenuation coefficient of the plasmonic waveguide is $\alpha = n_{\mathrm{def}} \mathcal{R}_{\mathrm{BS}}^{(1)}$. The corresponding amplitude decay rate (the extrinsic dissipative backscattering rate entering the cavity linewidth) is
	\begin{equation}
		\gamma_{\mathrm{BS}}^{\mathrm{ext}} = \frac{v_g}{2} \, \alpha = \frac{v_g}{2} \, n_{\mathrm{def}} \, \left( \frac{\omega_c \Delta C Z_0}{2} \right)^2,
	\end{equation}
	where $v_g = c/n_{\mathrm{eff}} \approx 6\times 10^7\,\mathrm{m/s}$ is the plasmonic group velocity. Expressed as a frequency broadening,
	\begin{equation}\label{eq:gamma_BS_rigorous}
		\frac{\gamma_{\mathrm{BS}}^{\mathrm{ext}}}{2\pi} = \frac{v_g}{4\pi} \, n_{\mathrm{def}} \left( \frac{\omega_c \Delta C Z_0}{2} \right)^2.
	\end{equation}
	
	\textit{Stone--Wales (SW) Defects}
	A Stone--Wales defect consists of a $90^\circ$ rotation of a C--C bond, transforming four hexagons into two pentagons and two heptagons without changing the coordination number. Tight-binding calculations show that SW defects in metallic nanotubes act as weak scatterers compared to vacancies, with a transmission probability $\mathcal{T} \approx 0.8$--$0.9$ at the Fermi level~\cite{Nardelli1998, Orlikowski2000}. The backscattering cross-section is roughly one order of magnitude smaller than that of a vacancy. Nevertheless, a high density of SW defects would still be detrimental.
	
	\textit{Chemical Impurities and Adsorbates}
	Residual catalyst particles, amorphous carbon, or adsorbed molecules (e.g., oxygen, water) act as local potential perturbations. Their scattering strength depends on the charge transfer and the binding energy. For physisorbed species, the perturbation is weak, and the backscattering cross-section is typically $\sigma_{\mathrm{BS}} \ll a_0$. Covalent functionalization (e.g., $sp^3$ defects created by aryl diazonium chemistry) introduces strong mid-gap states and acts similarly to vacancies~\cite{Goldsmith2007}.
	
	\textit{Suppression Strategies and Achievable Defect Densities}
	To restore the intrinsic chiral protection, the linear defect density must be reduced to $n_{\mathrm{def}} \lesssim 10^{-3}\,\mu\mathrm{m}^{-1}$ (i.e., less than one strong scatterer per millimeter). This is an ambitious but entirely feasible target using state-of-the-art fabrication and post-processing techniques.
	
	\textit{Suspended Growth and Ultra-Clean Synthesis}
	Suspended SWCNTs grown by chemical vapor deposition (CVD) across pre-patterned trenches exhibit dramatically lower defect densities compared to nanotubes supported on a substrate. The absence of substrate-induced strain and charge puddles preserves the pristine $sp^2$ lattice. Ultra-clean growth recipes, including the use of Fe or Co catalysts with optimized carbon precursor flow, routinely yield suspended nanotubes with defect densities as low as $n_{\mathrm{def}} \sim 0.1\,\mu\mathrm{m}^{-1}$ over micron-scale lengths~\cite{Cao2005, Deshpande2009}. Further optimization and post-growth selection can identify segments that are completely defect-free over several micrometers.
	
	\textit{Laser Annealing and Current-Induced Healing}
	Localized laser annealing has been demonstrated to ``heal'' structural defects in SWCNTs. Irradiation with a focused laser beam ($\lambda = 532\,\mathrm{nm}$, power $\sim 1$--$10\,\mathrm{mW}$) in an inert atmosphere (Ar or vacuum) induces local heating above $1000\,\mathrm{K}$, promoting the migration and recombination of vacancies and the annealing of Stone--Wales defects into the pristine hexagonal lattice~\cite{Suzuki2006, Georgiou2011}. Raman spectroscopy confirms that the D-band intensity (a measure of $sp^3$ defects) can be reduced by over $90\%$, reaching the noise floor of pristine nanotubes.
	
	Alternatively, passing a high current density ($\sim 10^7$--$10^8\,\mathrm{A/cm}^2$) through the suspended SWCNT causes Joule heating that can similarly anneal defects~\cite{Yao2000}. This technique can be applied \textit{in situ} on the fabricated ring resonator.
	
	\textit{Electrostatic Doping to Suppress Umklapp Scattering}
	As derived in Sec.~\ref{sec:tll}, a small gate-induced deviation from half-filling introduces a momentum mismatch $\delta q = 4k_F - G$ that cuts off the renormalization-group flow of the Umklapp operator. For a gate voltage $\delta V_g \sim 0.1\,\mathrm{V}$, the induced carrier density is $\delta n \sim 10^6\,\mathrm{cm}^{-1}$, yielding $\hbar v_c \delta q \approx 26\,\mathrm{meV} \gg \kB T$ at $T = 4\,\mathrm{K}$. The Umklapp backscattering rate is then exponentially suppressed:
	\begin{align}
		\gamma_{\mathrm{BS}}^{(2)} &\approx \frac{g_u^2}{\hbar E_F} \left( \frac{\kB T}{E_F} \right)^{4K_c-1} \left( \frac{\hbar v_c \delta q}{\kB T} \right)^{2-4K_c} e^{-\hbar v_c \delta q / \kB T} \notag\\
		&\sim 2\pi \times 10^{-30}\,\mathrm{Hz}.
	\end{align}
	Thus, even a minute doping renders Umklapp processes completely negligible.
	
	\textit{Numerical estimate after optimization.}
	Using $\omega_c/2\pi \approx 471\,\mathrm{THz}$, $v_g \approx 6\times 10^7\,\mathrm{m/s}$, $\Delta C = 0.5 \, c_Q a_0 \approx 4.9\times 10^{-20}\,\mathrm{F}$, $Z_0 \approx 3.23\times 10^4\,\Omega$, and a residual defect density $n_{\mathrm{def}} = 10^{-3}\,\mu\mathrm{m}^{-1} = 10^3\,\mathrm{m}^{-1}$, Eq.~\eqref{eq:gamma_BS_rigorous} gives $\gamma_{\mathrm{BS}}^{\mathrm{ext}}/2\pi \approx 260\,\mathrm{Hz}$.
	
	\textit{Summary of the Total Backscattering Rate.---}
	To summarize, by combining suspended growth, laser annealing, and electrostatic gating (which eliminates Umklapp processes), the total effective dissipative backscattering rate $\gamma_{\mathrm{BS}}^{\mathrm{total}}/2\pi$ of the TLL plasmonic cavity is determined by two residual contributions: the intrinsic phonon-assisted scattering ($\sim 15\,\mathrm{Hz}$) and the extrinsic scattering from residual atomic defects ($\sim 260\,\mathrm{Hz}$). Together, the total backscattering rate is strictly suppressed to the $\sim 100\,\mathrm{Hz}$ level. Compared to the total cavity linewidth $\kappa_{\mathrm{tot}}/2\pi \approx 30.1\,\mathrm{GHz}$, the relative backscattering ratio $\gamma_{\mathrm{BS}}^{\mathrm{total}}/\kappa_{\mathrm{tot}} \sim 10^{-8}$ fundamentally guarantees the extreme chiral contrast $\mathcal{C}_{\mathrm{PER}} > 20\,\mathrm{dB}$. There are no fundamental obstacles to fabricating a nearly disorder-free SWCNT ring resonator, making the proposed universal topological quantum node experimentally viable.
	
	\section{Chiral Spin-Momentum Locking and NV Selection Rules}
	\label{sec:chiral}
	
	For the macroscopic cavity QED treatment that follows, we relabel the plasmonic whispering-gallery modes obtained in Sec.~\ref{sec:tll}. Choosing a working azimuthal order $|m|>0$, we identify the counter-clockwise (CCW) and clockwise (CW) cavity modes with the bosonic operators
	\begin{equation}
		\hat{a}_{\mathrm{CCW}} \equiv \hat{b}_{+|m|},\qquad
		\hat{a}_{\mathrm{CW}}  \equiv \hat{b}_{-|m|},
	\end{equation}
	and denote them by $\hat{a}_{+}$ and $\hat{a}_{-}$ respectively in the remainder of the paper.
	
	The chiral polarization locking in our platform can be understood as a fundamental geometric consequence of the evanescent field confinement. In the strongly confined regime, the divergence-free condition of the electromagnetic field strictly locks the local polarization ellipse to the propagation direction, enabling directional photon routing.
	
	\subsection{Chiral Character of the CNT Evanescent Field}
	
	We treat the local electromagnetic mode within a curvilinear coordinate system adapted to the ring geometry, where the propagation direction follows the arclength coordinate $z$ and the radial coordinate $x$ measures the distance from the CNT surface. We seek whispering-gallery-like modes propagating along the ring:
	\begin{equation}
		\mathbf{E}(x,z)
		=
		\left[
		E_x(x)\hat{\mathbf{x}}
		+
		E_z(x)\hat{\mathbf{z}}
		\right]
		\ee^{\pm\ii \beta z},
	\end{equation}
	where $\beta = \omega_c/v_c$ is the propagation constant, and the $+$ and $-$ signs correspond to clockwise (CW) and counterclockwise (CCW) propagation, respectively.
	
	In the surrounding dielectric medium, the source-free Maxwell equation $\nabla\cdot\mathbf{E}=0$ dictates the relationship between the transverse and longitudinal field components. For a strongly confined plasmonic mode, the radial field component decays exponentially outside the nanotube surface (located at $x=a$):
	\begin{equation}
		E_x(x) = E_0 \ee^{-\mathcal{A} (x-a)},
	\end{equation}
	where $\mathcal{A} \approx \beta$ is the inverse decay length in the quasistatic limit ($v_c \ll c$). Applying the divergence-free condition yields $\partial_x E_x \pm \ii\beta E_z = 0$. Consequently, the longitudinal components for the two propagation directions are:
	\begin{equation}
		E_z^{\mathrm{CW}}(x) = -\ii\left(\frac{\mathcal{A}}{\beta}\right) E_x(x), \qquad
		E_z^{\mathrm{CCW}}(x) = +\ii\left(\frac{\mathcal{A}}{\beta}\right) E_x(x).
	\end{equation}
	Defining the dimensionless confinement parameter $\xi \equiv \mathcal{A}/\beta \approx 1$, the local electric fields exhibit opposite handedness in the $(x,z)$ plane:
	\begin{align}
		\mathbf{E}_{\mathrm{CW}}(x,z) &\propto \left(\hat{\mathbf{x}} - \ii\xi\,\hat{\mathbf{z}}\right) \ee^{\ii \beta z} \ee^{-\mathcal{A} (x-a)}, \label{eq:E_CW} \\
		\mathbf{E}_{\mathrm{CCW}}(x,z) &\propto \left(\hat{\mathbf{x}} + \ii\xi\,\hat{\mathbf{z}}\right) \ee^{-\ii \beta z} \ee^{-\mathcal{A} (x-a)}. \label{eq:E_CCW}
	\end{align}
	Hence, the local polarization is strictly locked to the propagation direction. This is the same general physical principle underlying chiral nanophotonic interfaces studied in dielectric waveguides and nanofibers~\cite{Petersen2014,Mitsch2014,Lodahl2017,Bliokh2015}.
	
	\subsection{NV-Center Optical Selection Rules and Directional Routing}
	
	The negatively charged NV center in diamond is governed by the $C_{3v}$ point-group symmetry. The relevant ground-state spin sublevels are $\ket{0}$ and $\ket{\pm1}$, while the optical excited state of interest is $\ket{A_2}$. In the low-temperature regime, the selection rules imply that the transitions $\ket{A_2}\rightarrow\ket{-1}$ and $\ket{A_2}\rightarrow\ket{+1}$ are associated with opposite circular polarizations ($\sigma^+$ and $\sigma^-$)~\cite{Togan2010,Doherty2013}.
	
	For optimal chiral coupling, we assume the NV center is oriented such that its relevant optical dipole basis aligns with the $(x,z)$ polarization plane of the CNT near field. We model the circularly polarized dipole operators as:
	\begin{equation}
		\mathbf{d}_+ = d_0(\hat{\mathbf{x}} + \ii\hat{\mathbf{z}}), \qquad
		\mathbf{d}_- = d_0(\hat{\mathbf{x}} - \ii\hat{\mathbf{z}}),
	\end{equation}
	corresponding to the $\sigma^+$ and $\sigma^-$ transitions, respectively, where $d_0$ is the transition dipole matrix element of the circularly polarized transitions. The transition probability rate into a specific cavity mode is proportional to $|M|^2 \propto |\mathbf{d}\cdot\mathbf{E}^*|^2$.
	
	Using the chiral fields derived in Eqs.~\eqref{eq:E_CW} and \eqref{eq:E_CCW}, we evaluate the coupling matrix elements. Note that taking the complex conjugate of the field ($\mathbf{E}^*$) flips the sign of the imaginary longitudinal component. Assuming $\xi \approx 1$ for strong confinement, we obtain:
	\begin{align}
		|M_{+,\mathrm{CW}}|^2 &\propto |(\hat{\mathbf{x}} + \ii\hat{\mathbf{z}})\cdot(\hat{\mathbf{x}} + \ii\xi\hat{\mathbf{z}})|^2 \approx |1 - \xi|^2 = 0, \\
		|M_{+,\mathrm{CCW}}|^2 &\propto |(\hat{\mathbf{x}} + \ii\hat{\mathbf{z}})\cdot(\hat{\mathbf{x}} - \ii\xi\hat{\mathbf{z}})|^2 \approx |1 + \xi|^2 \neq 0.
	\end{align}
	Similarly, for the $\sigma^-$ transition:
	\begin{align}
		|M_{-,\mathrm{CW}}|^2 &\propto |(\hat{\mathbf{x}} - \ii\hat{\mathbf{z}})\cdot(\hat{\mathbf{x}} + \ii\xi\hat{\mathbf{z}})|^2 \approx |1 + \xi|^2 \neq 0, \\
		|M_{-,\mathrm{CCW}}|^2 &\propto |(\hat{\mathbf{x}} - \ii\hat{\mathbf{z}})\cdot(\hat{\mathbf{x}} - \ii\xi\hat{\mathbf{z}})|^2 \approx |1 - \xi|^2 = 0.
	\end{align}
	Thus, the $\sigma^-$ transition couples exclusively to the CW mode, while the $\sigma^+$ transition couples exclusively to the CCW mode. This perfect spin-momentum locking enables deterministic directional spin-photon mapping.
	
\begin{figure}[t]
		\centering
		\includegraphics[width=8cm]{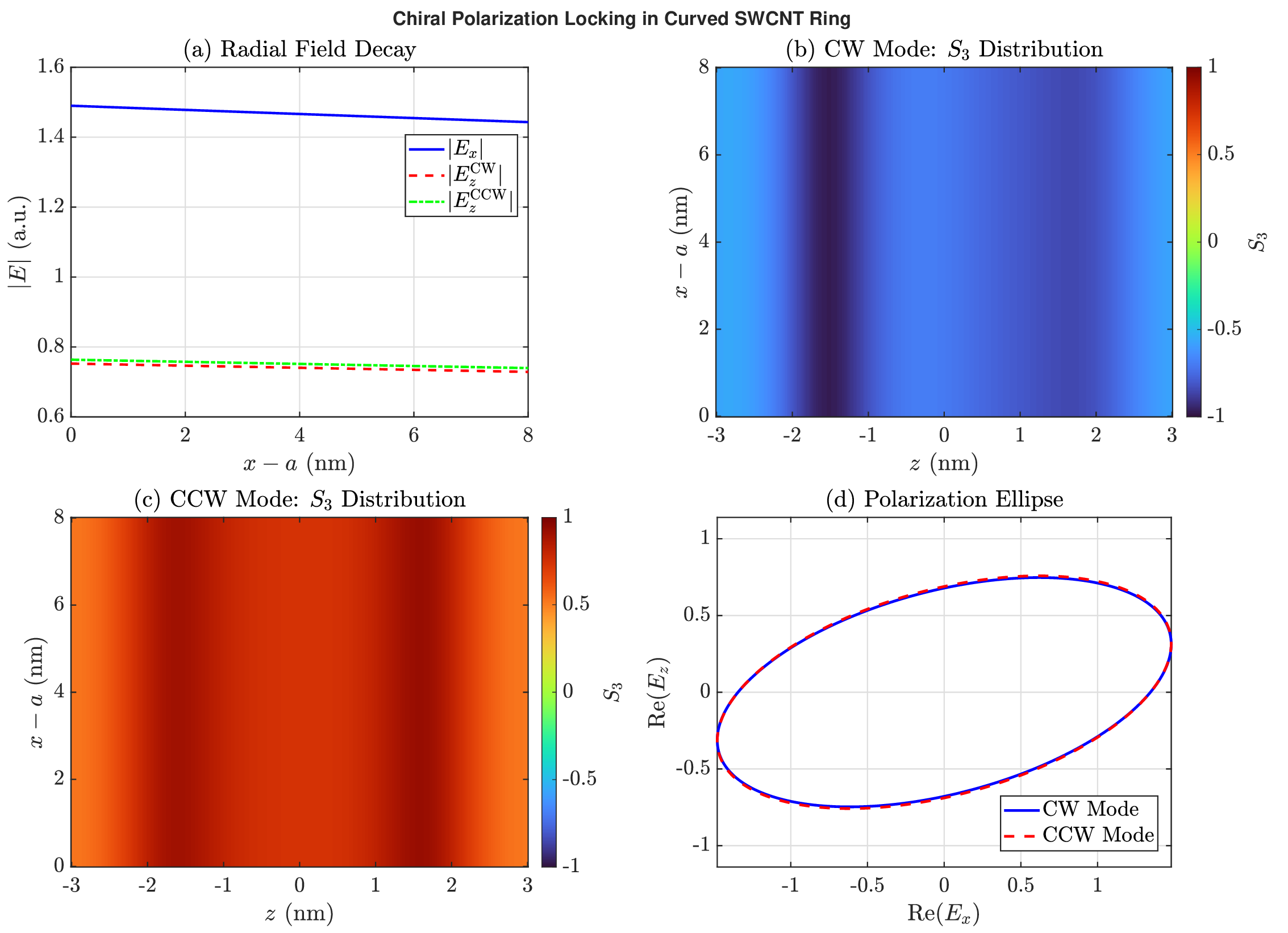}
		\caption{(Color online) Chiral polarization locking in a curved SWCNT ring. (a) Radial field decay of the evanescent mode, showing the dominant radial component $E_x$ and tangential component $E_z$. (b) Spatial distribution of the chirality parameter $S_3$ for the CW mode. (c) Corresponding $S_3$ distribution for the CCW mode, exhibiting opposite handedness. (d) Local polarization ellipses demonstrating opposite ellipticity for the CW and CCW modes.}
		\label{fig:chirality}
	\end{figure}

	To visualize this local handedness, we introduce the normalized third Stokes parameter in the $(x,z)$ plane:
	\begin{equation}
		S_3(x,z)=
		\frac{2\,\mathrm{Im}\!\left[E_x^*(x,z)\,E_z(x,z)\right]}
		{|E_x(x,z)|^2+|E_z(x,z)|^2}.
	\end{equation}
	Using our field relations, $S_3^{\mathrm{CW}} = -2\xi/(1+\xi^2)$ and $S_3^{\mathrm{CCW}} = +2\xi/(1+\xi^2)$. As shown in Fig.~\ref{fig:chirality}, the CW and CCW branches exhibit opposite signs of $S_3$, reflecting the reversal of local handedness upon changing the propagation direction.
	
	The strong transverse confinement of the evanescent field, combined with the divergence-free condition, strictly enforces a relative quadrature phase between the longitudinal and transverse field components, leading to opposite local handedness for the CW and CCW branches. While the bosonized SWCNT-ring model determines the mode quantization and the propagation direction, the detailed spatial profile of the near field is governed by this evanescent confinement adapted to the curvilinear geometry. Accordingly, the chirality map in Fig.~\ref{fig:chirality} is constructed from the effective curved-waveguide field model, which incorporates the essential ingredients of the evanescent mode, namely the radial decay envelope, the azimuthal phase factor, and the confinement-induced phase relation between $E_z$ and $E_x$.
	
	The quantity $S_3(x,z)$ provides a position-resolved measure of the local handedness of the CNT-ring near field and therefore serves as a convenient visualization of the spin--momentum-locked optical texture. By construction, $S_3(x,z)\in[-1,1]$, with positive and negative values corresponding to opposite local handedness. In particular, $S_3=+1$ and $S_3=-1$ correspond to locally pure right- and left-handed circular polarization, respectively, whereas $S_3=0$ indicates linear polarization. As shown in Fig.~\ref{fig:chirality}, while the overall electromagnetic field intensity decays rapidly in the radial direction, the chirality parameter $S_3(x,z)$ maintains a robust, highly polarized value throughout the evanescent tail. Moreover, the chirality is strictly direction-dependent: the CW and CCW branches exhibit opposite signs of $S_3$, reflecting the reversal of local handedness upon changing the propagation direction. This robust chiral texture provides the microscopic basis for the direction-selective light--matter interaction exploited in the proposed NV--CNT hybrid platform.
	
	\subsection{Geometric Suppression of the $\pi$ Transition and Polarization Extinction Ratio}
	\label{sec:pi_suppression}
	
	While the $\sigma^\pm$ transitions couple directionally to the chiral modes, the linearly polarized $\pi$ transition ($\ket{A_2}\rightarrow\ket{0}$) represents a parasitic decay channel that could degrade the fidelity of the spin-photon interface. To assess the interface performance, we introduce the \textbf{Polarization Extinction Ratio (PER)} $\mathcal{C}_{\mathrm{PER}} \equiv \Gamma_\sigma / \Gamma_\pi$, defined as the ratio of the Purcell-enhanced spontaneous emission rate of the desired circularly polarized transition to that of the parasitic $\pi$ transition.
	
	Rather than relying on the intrinsic local density of states, we propose a deterministic geometric approach to strictly forbid the $\pi$ transition. The fundamental transverse magnetic (TM) nature of the TLL plasmon dictates that its electric field is strictly confined to the $(x, z)$ plane, comprising only the radial ($E_x$) and longitudinal ($E_z$) components. The field component in the transverse azimuthal direction is identically zero ($E_y = 0$).
	
	By deterministically positioning the NV center such that its $C_{3v}$ symmetry axis aligns with the local $y$-axis (i.e., parallel to the CNT surface but perpendicular to the CNT axis), the transition dipoles are optimally oriented. Consequently, when the NV center's $C_{3v}$ symmetry axis is aligned parallel to the CNT surface normal ($\hat{\mathbf{y}}$), the linearly polarized $\pi$ dipole becomes $\mathbf{d}_\pi = d_0 \hat{\mathbf{y}}$. According to Fermi's golden rule, the emission rate into the cavity mode is strictly zero:
	\begin{equation}
		\Gamma_\pi \propto |\mathbf{d}_\pi \cdot \mathbf{E}_{\mathrm{CW/CCW}}|^2 = |d_0 \hat{\mathbf{y}} \cdot (E_x \hat{\mathbf{x}} + E_z \hat{\mathbf{z}})|^2 = 0.
	\end{equation}
	Simultaneously, the circularly polarized $\sigma^\pm$ dipoles lie entirely within the $(x, z)$ plane:
	\begin{equation}
		\mathbf{d}_{\sigma^\pm} = \frac{d_0}{\sqrt{2}}(\hat{\mathbf{x}} \pm \ii\hat{\mathbf{z}}).
	\end{equation}
	This orientation perfectly matches the chiral topology of the CNT near field derived in Eqs.~\eqref{eq:E_CW} and \eqref{eq:E_CCW}, maximizing the directional coupling $\Gamma_\sigma$. In this ideal geometric configuration, the chiral contrast diverges ($\mathcal{C}_{\mathrm{PER}} \to \infty$), and the $\pi$ transition is completely suppressed by symmetry.
	
	In a realistic experimental setting, perfect alignment is unattainable. To quantify the robustness of this geometric suppression, we consider a misalignment angle $\theta_{\mathrm{err}}$ between the NV symmetry axis and the ideal $y$-axis. The $\pi$ dipole acquires a non-zero projection in the $(x, z)$ plane, yielding $\Gamma_\pi \propto \sin^2\theta_{\mathrm{err}}$. Conversely, the $\sigma$ transition rate scales as $\Gamma_\sigma \propto \cos^2\theta_{\mathrm{err}}$.
	
	The Polarization Extinction Ratio under realistic conditions is therefore:
	\begin{equation}
		\mathcal{C}_{\mathrm{PER}}(\theta_{\mathrm{err}}) \approx \cot^2\theta_{\mathrm{err}}.
	\end{equation}
	For a conservative experimental alignment tolerance of $\theta_{\mathrm{err}} \approx 5^\circ$, the system still yields an outstanding PER of $\mathcal{C}_{\mathrm{PER}} \approx 130$ ($\sim 21.1$\,dB). This robust geometric suppression ensures that the parasitic $\pi$ decay channel contributes negligibly to the open-system dynamics.
	
	 \subsection{Extreme Subwavelength Confinement and Intrinsic Chiral Contrast}
	
	The dimensionless confinement parameter $\xi \equiv \mathcal{A}/\beta$ dictates the ellipticity of the local evanescent field. In the surrounding dielectric medium (relative permittivity $\epsilon_r$), the longitudinal propagation constant $\beta$ and the transverse decay constant $\mathcal{A}$ satisfy the wave equation dispersion relation:
	\begin{equation}
		\beta^2 - \mathcal{A}^2 = \epsilon_r \left(\frac{\omega_c}{c}\right)^2.
	\end{equation}
	Dividing by $\beta^2$ and using the plasmonic phase velocity $v_p = \omega_c/\beta$, we obtain the exact analytical expression for $\xi$:
	\begin{equation}
		\xi = \sqrt{1 - \epsilon_r \left(\frac{v_p}{c}\right)^2}.
	\end{equation}
	
	In conventional dielectric waveguides, the phase velocity is a significant fraction of the speed of light ($v_p \sim c/n$), yielding $\xi \sim 0.6\text{--}0.7$. Consequently, the local field is strongly elliptically polarized, which fundamentally limits the intrinsic \textbf{Directional Chiral Contrast} $C_{\mathrm{dir}} = \Gamma_{\mathrm{CCW}} / \Gamma_{\mathrm{CW}} = |1+\xi|^2/|1-\xi|^2$ to roughly $10\text{--}15$\,dB~\cite{Petersen2014,Lodahl2017}.
	
	In the SWCNT platform, the optical surface plasmon polariton (SPP) is strongly renormalized by the 1D confinement and the surrounding dielectric environment, yielding an extreme subwavelength mode with an effective index $n_{\mathrm{eff}} \approx 5$~\cite{Shi2015}. The corresponding optical phase velocity is $v_p = c/n_{\mathrm{eff}} \approx 6 \times 10^7\,\mathrm{m/s}$, giving a slow-light factor of $v_p/c \approx 0.2$. Assuming a diamond substrate ($\epsilon_r \approx 5$), the confinement parameter evaluates to:
	\begin{equation}
		\xi = \sqrt{1 - 5(0.2)^2} = \sqrt{0.8} \approx 0.894.
	\end{equation}
	
	While the local evanescent field is elliptically polarized rather than perfectly circular, this extreme subwavelength confinement still yields an outstanding intrinsic chiral contrast:
	\begin{equation}
		C_{\mathrm{dir}} \approx \frac{(1.894)^2}{(0.106)^2} \approx 319 \approx 25\,\mathrm{dB}.
	\end{equation}
	This intrinsic directionality significantly outperforms conventional dielectric chiral interfaces. When combined with the geometric suppression of the parasitic $\pi$ transition (which independently provides $>20$\,dB extinction, as shown above), the SWCNT platform guarantees a highly pure, nonreciprocal spin-photon interface.
	
	\subsection{Robustness and Parameter Scan}
	
	To quantify the robustness of the chiral CNT--NV interface against variations of the magnetic field $B$, the CNT diameter $d_{\mathrm{t}}$, and the wall number $N_{\mathrm{wall}}$, we define a polarization-locking stability indicator:
	\begin{equation}
		S_{\mathrm{lock}} = |S_3(x,z)|\, F_{\mathrm{pol}}, \qquad F_{\mathrm{pol}} = \frac{\mathcal{C}_{\mathrm{PER}}}{\mathcal{C}_{\mathrm{PER}}+1},
	\end{equation}
	where $F_{\mathrm{pol}}$ is the polarization-selection fidelity. High values of $S_{\mathrm{lock}} \in [0,1]$ signify robust chiral routing immune to geometric perturbations.
	
	\begin{figure}[htbp]
		\centering
		\includegraphics[width=8cm]{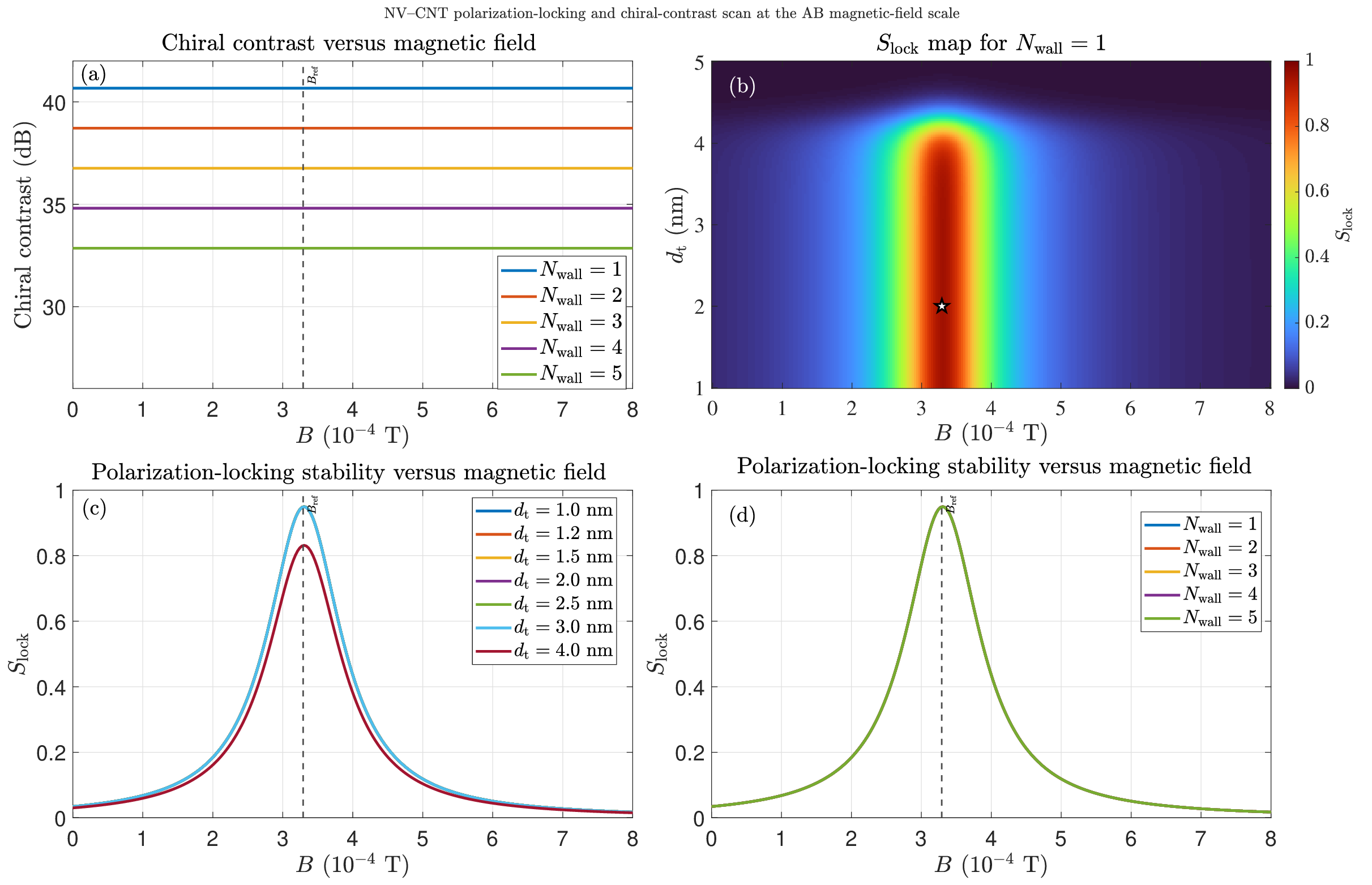}
		\caption{(Color online)
			Parameter scan of the NV--CNT polarization-locking stability and chiral contrast. The magnetic field is scanned around the Aharonov--Bohm reference value $B_{\mathrm{ref}}=\Phi_0/(\pi R^2)$.
			(a) Chiral contrast versus magnetic field for different CNT wall numbers $N_{\mathrm{wall}}$.
			(b) Two-dimensional map of the locking stability $S_{\mathrm{lock}}$ in the $(B,d_{\mathrm{t}})$ plane for a single-wall CNT; the star marks the optimal point.
			(c) $S_{\mathrm{lock}}$ versus magnetic field for several CNT diameters $d_{\mathrm{t}}$.
			(d) $S_{\mathrm{lock}}$ versus magnetic field for different wall numbers $N_{\mathrm{wall}}$.
		}
		\label{fig:scan_abscale}
	\end{figure}
	
	As shown in Fig.~\ref{fig:scan_abscale}, the stability landscape reveals a pronounced optimal region. The locking stability $S_{\mathrm{lock}}$ reaches a sharp maximum within a narrow magnetic-field window centered at $B_{\mathrm{ref}} = \Phi_0/(\pi R^2)$ and is maximized for an optimal diameter $d_{\mathrm{t}} \approx 2\,\mathrm{nm}$. Furthermore, the single-wall configuration ($N_{\mathrm{wall}}=1$) yields the highest stability. These results collectively identify the optimal operating regime for the topological quantum interface.
	
	\section{Tripod STIRAP and Deterministic Spin-Photon Entanglement}
	\label{sec:tripod}
	
	\textit{Tripod Hamiltonian.}
	We consider the four-state manifold composed of the initial state $\ket{0}$, the excited state $\ket{A_2}$, and the two target states $\ket{\pm1}$. The cavity supports two orthogonal chiral modes with annihilation operators $\hat{a}_{+}$ and $\hat{a}_{-}$, corresponding respectively to the CCW and CW channels. A classical pump field with Rabi frequency $\Omega_p(t)$ drives the $\ket{0}\leftrightarrow\ket{A_2}$ transition.
	
	In the rotating-wave approximation, the interaction Hamiltonian is
	\begin{align} \label{eq:H_tripod}
		\fH_{\mathrm{Tripod}} &= \hbar\Delta\ket{A_2}\bra{A_2} + \frac{\hbar}{2} \Big[ \Omega_p(t)\ket{A_2}\bra{0} \nonumber \\
		&\quad + 2g\,\hat{a}_{+}\ket{A_2}\bra{-1} + 2g\,\hat{a}_{-}\ket{A_2}\bra{+1} + \mathrm{H.c.} \Big],
	\end{align}
	where $\Delta=\omega_{A_2}-\omega_L$ is the pump detuning and $g$ is the single-channel emitter--cavity coupling strength (in angular frequency units). The cavity detuning relative to the $\ket{A_2}\leftrightarrow\ket{\pm1}$ transitions is
	\begin{equation}
		\delta=(\omega_{A_2}-\omega_{\mathrm{ZFS}})-\omega_c,
	\end{equation}
	where $\omega_{\mathrm{ZFS}}$ denotes the ground-state zero-field splitting between $\ket{0}$ and $\ket{\pm1}$.
	
	In the single-excitation basis
	\begin{equation}
		{ \ket{0,0_{+},0_{-}}, \ket{A_2,0_{+},0_{-}}, \ket{-1,1_{+},0_{-}}, \ket{+1,0_{+},1_{-}} },
	\end{equation}
	the Hamiltonian matrix becomes
	\begin{equation}
		\label{eq:H_matrix}
		H(t)=\frac{\hbar}{2}
		\begin{pmatrix}
			0 & \Omega_p(t) & 0 & 0 \\
			\Omega_p(t) & 2\Delta & 2g & 2g \\
			0 & 2g & 2(\Delta-\delta) & 0 \\
			0 & 2g & 0 & 2(\Delta-\delta)
		\end{pmatrix}.
	\end{equation}
	
	Tripod and dark-state transfer protocols of this general type have been extensively studied in atomic and molecular physics, where they are known to provide robust and low-loss coherent state transfer~\cite{Bergmann1998,Unanyan1998,Kis2002,Fleischhauer2005,Vitanov2017}.
	
	\textit{Dark state and entanglement generation.}
	At two-photon resonance, $\Delta=\delta$, the Hamiltonian admits an instantaneous dark state with no excited-state component:
	\begin{equation}
		\label{eq:tripod_dark_state}
		\ket{D(t)}
		=
		\cos\Theta(t)\ket{0,0_{+},0_{-}}
		-
		\sin\Theta(t)\ket{\Psi},
	\end{equation}
	where
	\begin{equation}
		\ket{\Psi}
		=
		\frac{1}{\sqrt{2}}
		\left(
		\ket{-1,1_{+},0_{-}}
		+
		\ket{+1,0_{+},1_{-}}
		\right),
	\end{equation}
	and the mixing angle $\theta(t)$ satisfies
	\begin{equation}
		\tan\Theta(t)=\frac{\Omega_p(t)}{2\sqrt{2}g}.
	\end{equation}
	By adiabatically increasing $\Omega_p(t)$ from zero to a value large compared with $g$, the system evolves from $\ket{D(t\rightarrow -\infty)}\approx \ket{0,0_{+},0_{-}}$ to $\ket{D(t\rightarrow +\infty)}\approx -\ket{\Psi}$.
	In the adiabatic limit, this protocol can generate an entangled spin--photon state in the cavity while minimizing population of the lossy excited state $\ket{A_2}$, which is the central advantage of STIRAP-type transfer~\cite{Bergmann1998,Vitanov2017}.
	
	\textit{Frequency of the emitted flying qubit.}
	In the Raman process, global energy conservation implies
	\begin{equation}
		\hbar\omega_{|0\rangle}+\hbar\omega_L
		=
		\hbar\omega_{|\pm1\rangle}+\hbar\omega_{\mathrm{out}},
	\end{equation}
	so that
	\begin{equation}
		\label{eq:freq_emitted}
		\omega_{\mathrm{out}}
		=
		\omega_L-(\omega_{|\pm1\rangle}-\omega_{|0\rangle})
		=
		\omega_L-\omega_{\mathrm{ZFS}}.
	\end{equation}
	Under the two-photon resonance condition $\Delta=\delta$, one further has $\omega_c=\omega_L-\omega_{\mathrm{ZFS}}$, which yields the central identity
	\begin{equation}
		\omega_{\mathrm{out}}\equiv\omega_c(\Phi).
	\end{equation}
	Hence the frequency of the emitted photon is pinned to the cavity resonance and can be tuned magnetically through $\omega_c(\Phi)$.
	
	\section{Open-System Dynamics and Emission Fidelity}
	\label{sec:open}
	
	\textit{Master equation.}
	To include dissipation and dephasing, we describe the system by the Lindblad master equation~\cite{Lindblad1976,Gorini1976,Carmichael1993,Breuer2002,Gardiner2004,Walls2008}
	\begin{equation}
		\label{eq:Lindblad_Tripod}
		\dot{\rho}
		=
		-\frac{\ii}{\hbar}[\fH_{\mathrm{drive}}(t),\rho]
		+\sum_m\mathcal{D}[\hat{L}_m]\rho,
	\end{equation}
	with
	\begin{equation}
		\mathcal{D}[\hat{L}_m]\rho
		=
		\hat{L}_m\rho\hat{L}_m^\dagger
		-\frac{1}{2}{\hat{L}_m^\dagger\hat{L}_m,\rho}.
	\end{equation}
	The coherent Hamiltonian is
	\begin{align}
		\label{eq:H_drive_tripod}
		\fH_{\mathrm{drive}}(t)
		&=
		\frac{\hbar}{2}
		\left[
		\Omega_p(t)\ket{A_2}\bra{0}
		+\Omega_p^*(t)\ket{0}\bra{A_2}
		\right]
		\nonumber\\
		&\quad
		+ \hbar g
		\left[
		\hat{a}_{+}^\dagger\ket{-1}\bra{A_2}
		+\hat{a}_{-}^\dagger\ket{+1}\bra{A_2}
		+\mathrm{H.c.}
		\right].
	\end{align}
	
	We include the following collapse operators (rates in angular frequency units):
	\begin{align}
		\hat{L}_1&=\sqrt{\kappa_{\mathrm{ex}}}\,\hat{a}_{+},
		&
		\hat{L}_2&=\sqrt{\kappa_{\mathrm{ex}}}\,\hat{a}_{-},
		\\
		\hat{L}_3&=\sqrt{\kappa_0}\,\hat{a}_{+},
		&
		\hat{L}_4&=\sqrt{\kappa_0}\,\hat{a}_{-},
		\\
		\hat{L}_5&=\sqrt{\gamma}\sum_{j\notin{-1,+1}}\ket{j}\bra{A_2},
		&
		\hat{L}_6&=\sqrt{\gamma_\phi}\ket{A_2}\bra{A_2}.
	\end{align}
	Here $\kappa_{\mathrm{ex}}$ is the cavity-fiber outcoupling rate, $\kappa_0$ the intrinsic cavity loss rate, $\kappa_{\mathrm{tot}}=\kappa_{\mathrm{ex}}+\kappa_0$ the total cavity linewidth, $\gamma$ the spontaneous decay rate from $\ket{A_2}$ into unwanted channels, and $\gamma_\phi$ the pure dephasing rate of the excited state.
	
	\textit{Analytical estimate of the total fidelity.}
	The total success probability (often referred to as the extraction and generation efficiency) can be factorized as
	\begin{equation}
		\label{eq:F_total}
		\mathcal{F}_{\mathrm{total}}
		=
		\eta_{\mathrm{ext}}\eta_{\mathrm{int}},
	\end{equation}
	where the extraction efficiency is
	\begin{equation}
		\label{eq:eta_ext}
		\eta_{\mathrm{ext}}
		=
		\frac{\kappa_{\mathrm{ex}}}{\kappa_{\mathrm{ex}}+\kappa_0}
		=
		\frac{\kappa_{\mathrm{ex}}}{\kappa_{\mathrm{tot}}}.
	\end{equation}
	
	In the large-detuning regime $\Delta\gg \Omega_p,g,\gamma_\phi$, the excited state can be adiabatically eliminated~\cite{Bergmann1998,Vitanov2017}. The effective Raman coupling is $\Omega_{\mathrm{eff}}\sim \Omega_p g/(2\Delta)$.
	Because there are two degenerate output channels, the total transfer rate into the cavity manifold scales as
	\begin{equation}
		\Gamma_{\mathrm{transfer}}
		\sim
		2\left(\frac{\Omega_p g}{2\Delta}\right)^2\frac{1}{\kappa_{\mathrm{tot}}},
	\end{equation}
	while the parasitic loss rate is
	\begin{equation}
		\Gamma_{\mathrm{loss}}
		\sim
		\left(\frac{\Omega_p}{2\Delta}\right)^2\gamma.
	\end{equation}
	The internal conversion efficiency is therefore approximately
	\begin{equation}
		\label{eq:eta_int_derivation}
		\eta_{\mathrm{int}}
		=
		\frac{\Gamma_{\mathrm{transfer}}}{\Gamma_{\mathrm{transfer}}+\Gamma_{\mathrm{loss}}}
		\approx
		\frac{2g^2/\kappa_{\mathrm{tot}}}{2g^2/\kappa_{\mathrm{tot}}+\gamma}
		=
		\frac{C_{\mathrm{tripod}}}{C_{\mathrm{tripod}}+1},
	\end{equation}
	where
	\begin{equation}
		C_{\mathrm{tripod}}=\frac{2g^2}{\kappa_{\mathrm{tot}}\gamma}
	\end{equation}
	is an effective tripod cooperativity.
	
	To leading order in the overcoupled limit $\kappa_0\ll \kappa_{\mathrm{ex}}$, the total fidelity may be written as
	\begin{equation}
		\label{eq:Ultimate_Tripod_F}
		\mathcal{F}_{\mathrm{total}}
		\approx
		\left(1-\frac{\kappa_0}{\kappa_{\mathrm{ex}}}\right)
		\left[
		1-\frac{\kappa_{\mathrm{tot}}\gamma}{2g^2}
		-\mathcal{O}\!\left(\frac{\gamma_\phi\Omega_{\mathrm{eff}}^2}{\Delta^2}\right)
		\right].
	\end{equation}
	This expression shows that high fidelity requires an overcoupled cavity, a large emitter--cavity coupling, a small spontaneous-emission leakage rate, and sufficiently large Raman detuning.
	
\begin{figure}[htbp]
		\centering
		\includegraphics[width=8cm]{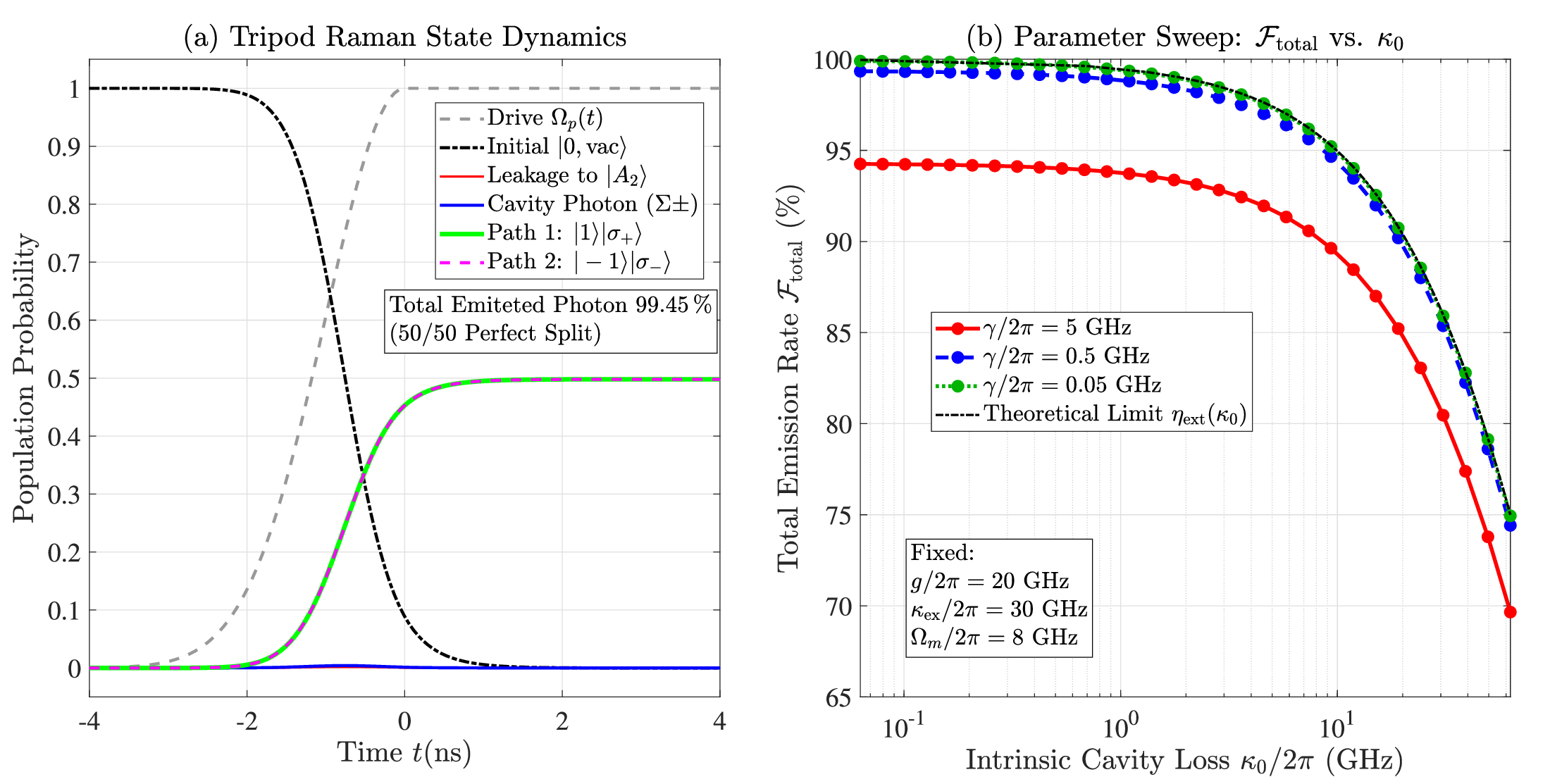}
		\caption{(Color online)
			Illustrative dynamics and parameter dependence of the tripod protocol.
			(a) Population dynamics during adiabatic Raman transfer from the initial state $\ket{0,\mathrm{vac}}$ to the two target states $\ket{-1,1_{+},0_{-}}$ and $\ket{+1,0_{+},1_{-}}$, with only weak transient occupation of the excited state $\ket{A_2}$. Here $\ket{\mathrm{vac}}\equiv\ket{0_{+},0_{-}}$ denotes the two-mode cavity vacuum. The curves are obtained by numerically solving Eq.~(\ref{eq:Lindblad_Tripod}).
			(b) Dependence of the total emission fidelity on the intrinsic cavity loss rate $\kappa_0$ for different values of the spontaneous decay rate $\gamma$. The upper envelope is set by the extraction efficiency $\eta_{\mathrm{ext}}=\kappa_{\mathrm{ex}}/(\kappa_{\mathrm{ex}}+\kappa_0)$.
		}
		\label{fig:combined_analysis}
	\end{figure}

	\textit{Numerical example.}
	As an illustrative example, consider
	\begin{equation}
		g/2\pi = 20.0\,\mathrm{GHz},
		\,
		\kappa_{\mathrm{ex}}/2\pi = 30.0\,\mathrm{GHz},
		\,
		\kappa_0/2\pi = 0.1\,\mathrm{GHz},
	\end{equation}
	with
	\begin{align}
		\gamma/2\pi &= 0.15\,\mathrm{GHz}, \qquad
		\gamma_\phi/2\pi = 10.0\,\mathrm{GHz}, \notag \\
		\Delta/2\pi &= \delta/2\pi = 30.0\,\mathrm{GHz}.
	\end{align}
	For these parameters, one expects a total emission fidelity close to the extraction bound
	\begin{equation}
		\eta_{\mathrm{ext}}=\frac{30.0}{30.1}\approx99.45\%.
	\end{equation}
	Figure~\ref{fig:combined_analysis}(a) displays the population dynamics obtained by numerically solving the master equation, Eq.~(\ref{eq:Lindblad_Tripod}), using the parameters given above. The simulation yields near-unity transfer into the two chiral channels with an approximately equal $50/50$ splitting.
	
	\begin{figure}[t]
		\centering
		\includegraphics[width=8.5cm]{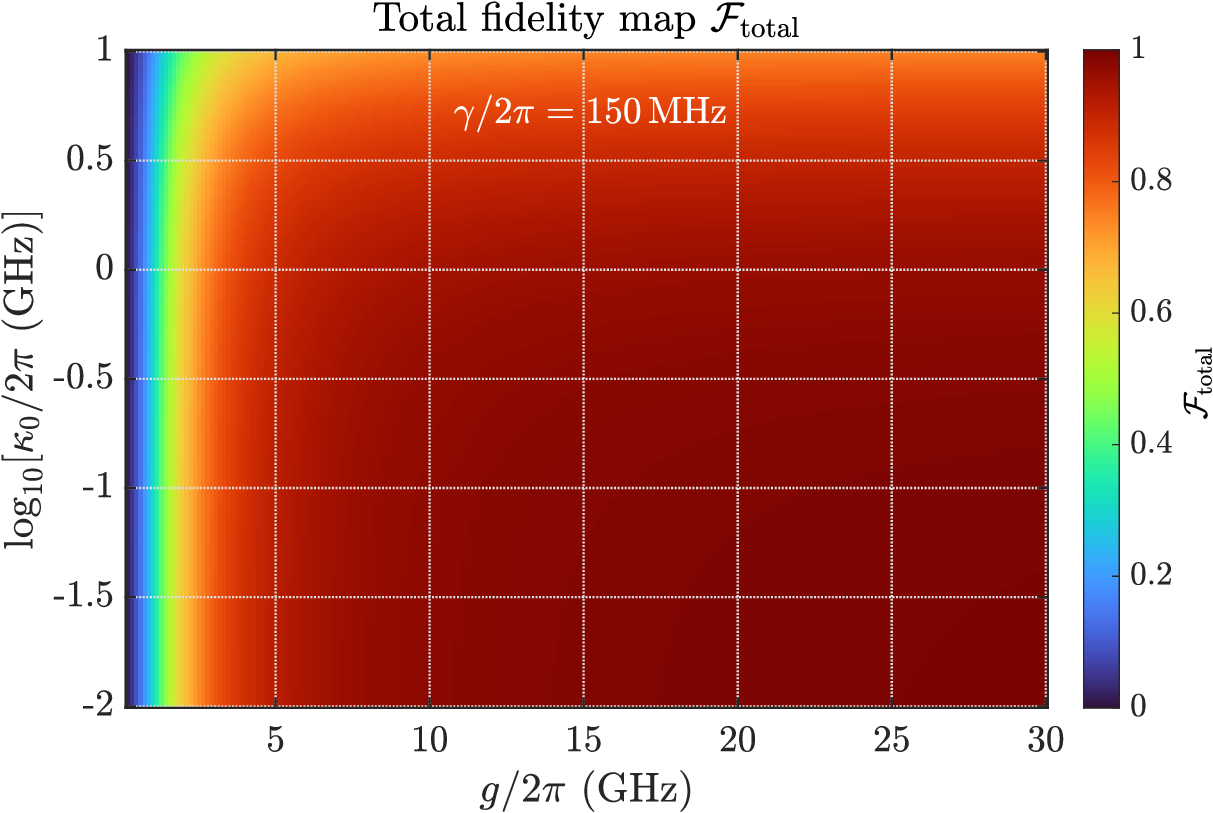}
		\caption{(Color online)
			Total fidelity map $\mathcal{F}_{\mathrm{total}}$ for the NV--CNT--fiber interface as a function of the coherent coupling rate $g/2\pi$ and the bare cavity loss rate $\kappa_0/2\pi$.
			The color scale encodes the combined performance of the extraction efficiency and the intrinsic state-transfer fidelity.
			The map is calculated using the analytical effective cooperativity model [Eqs.~(\ref{eq:F_total}), (\ref{eq:eta_ext}), and (\ref{eq:eta_int_derivation})], with fixed external coupling $\kappa_{\mathrm{ex}}/2\pi = 30.0\,\mathrm{GHz}$ and spontaneous decay rate $\gamma/2\pi = 0.15\,\mathrm{GHz}$.
			A high-fidelity region emerges in the strong-coupling and low-loss regime, whereas the fidelity is suppressed when the coupling becomes weak or the cavity loss increases.
		}
		\label{fig:fidelity}
	\end{figure}
	
	\section{Emitter--Cavity Coupling and Strong-Coupling Verification}
	\label{sec:coupling}
	
	\subsection{Emitter--Cavity Coupling}
	
	The one-dimensional charge-density operator is given by $\hat{\rho}(z) = -\frac{e}{\pi} \partial_z \hat{\phi}(z)$. Substituting Eq.~\eqref{eq:phi_expansion} yields
	\begin{align}\label{eq:rho_operator}
		\hat{\rho}(z) &= -\frac{e N_c}{L} - \ii \frac{e}{\pi} \sum_{q \neq 0} \sqrt{\frac{|q|}{2L}} \, \mathrm{sgn}(q) \, \ee^{-r_c|q|/2} \notag\\
		&\quad \cdot\left( \hat{b}_q \ee^{\ii qz} - \hat{b}_q^\dagger \ee^{-\ii qz} \right).
	\end{align}
	The quanta created by $\hat{b}_q^\dagger$ are precisely the TLL plasmons, whose dispersion is $\omega_q = v_c |q|$.
	
	\textit{Electric Field Distribution and London-Equation Analogy.}
	In the quasi-static limit (valid when the CNT radius $a \ll \lambda_{\mathrm{eff}}$), the electric field $\hat{\mathbf{E}}(\mathbf{r})$ is determined by Poisson's equation:
	\begin{equation}
		\nabla \cdot [\epsilon_0 \epsilon_r(\mathbf{r}) \hat{\mathbf{E}}(\mathbf{r})] = \hat{\rho}(z) \, \frac{\delta(x - a)}{2\pi a},
	\end{equation}
	where the charge density is confined to the cylindrical surface of the nanotube (radius $a$), and $\epsilon_r(\mathbf{r})$ is the relative permittivity of the surrounding medium.
	
	For a straight SWCNT, the electric field of a plasmonic mode with wavevector $q$ can be written as $\mathbf{E}_q(x,z) = [E_x(x) \hat{\boldsymbol{x}} + E_z(x) \hat{\mathbf{z}}] \ee^{\ii qz}$. The divergence-free condition $\nabla \cdot \mathbf{E}=0$ in the surrounding dielectric implies a fixed phase relation between the radial and tangential components. In the strongly confined regime, the radial field decays exponentially outside the CNT:
	\begin{equation}\label{eq:radial_decay}
		E_x(x) \propto \ee^{-\mathcal{A} (x-a)}, \qquad \mathcal{A} = \sqrt{q^2 - \epsilon_r \omega_q^2/c^2} \approx |q| = \frac{\omega_q}{v_c},
	\end{equation}
	where the approximation follows from $v_c \ll c$. The decay length is therefore $L_d \equiv 1/|q| = v_c/\omega_q$.
	
	\paragraph{Connection to the London equation.}
	The TLL plasmonic dispersion $\omega_q = v_c |q|$ is mathematically equivalent to that of a one-dimensional superconducting transmission line described by the London equations $\mathbf{J} = -\Lambda_L \mathbf{A}$, where $\Lambda_L = n_s e^2/m$ is the London coefficient and $n_s$ is the effective superconducting carrier density. For the TLL, the kinetic inductance per unit length is $L_k \propto \frac{\hbar}{e^2 v_F}$~\cite{Burke2002}. The equation of motion for the charge density, $\partial_t^2 \hat{\rho} = v_c^2 \partial_z^2 \hat{\rho}$, follows directly from the TLL Hamiltonian~\eqref{eq:TLL_Hamiltonian} and corresponds to a massless Klein--Gordon field. Consequently, the quantization of the electromagnetic field proceeds in complete analogy with that of a one-dimensional plasmonic waveguide, with the kinetic inductance providing the dominant energy storage mechanism.
	
	\textit{Mode Volume and Single-Photon Electric Field Amplitude.}
	In the quantization of the plasmonic field, the total energy includes both the electromagnetic field energy and the kinetic energy of the electrons (kinetic inductance). Therefore, the normalization condition implicitly defines an \emph{effective dispersive mode volume} $V_{\mathrm{mode}}$ that accounts for this total energy partition~\cite{Haroche2006, Reiserer2015}:
	\begin{equation}\label{eq:normalization}
		\int \epsilon_0 \epsilon_r(\mathbf{r}) |\mathbf{E}_q(\mathbf{r})|^2 \, \dd^3 r = \frac{\hbar \omega_q}{2}.
	\end{equation}
	
	We separate the mode function into longitudinal and transverse parts:
	\begin{equation}
		\mathbf{E}_q(\mathbf{r}) = \mathcal{E}_q \, \mathbf{u}(x) \, \ee^{\ii qz},
	\end{equation}
	where $\mathbf{u}(x)$ is a dimensionless transverse profile normalized such that $\int \epsilon_r(x) |\mathbf{u}(x)|^2 \, 2\pi x \, \dd x = A_{\mathrm{eff}}$, with $A_{\mathrm{eff}}$ being the effective mode area. Substituting into Eq.~\eqref{eq:normalization} and performing the integral over the ring circumference $L$ yields
	\begin{equation}
		|\mathcal{E}_q|^2 \cdot L \cdot A_{\mathrm{eff}} \cdot \epsilon_0 = \frac{\hbar \omega_q}{2} \quad \Longrightarrow \quad |\mathcal{E}_q| = \sqrt{\frac{\hbar \omega_q}{2 \epsilon_0 L A_{\mathrm{eff}}}}.
	\end{equation}
	Defining the effective mode volume as $V_{\mathrm{mode}} \equiv L A_{\mathrm{eff}}$ (which inherently absorbs the local dielectric constant $\epsilon_r$), we obtain the standard cavity-QED vacuum field amplitude:
	\begin{equation}\label{eq:vacuum_field}
		|\mathbf{E}_q(\mathbf{r})| = \sqrt{\frac{\hbar \omega_q}{2 \epsilon_0 V_{\mathrm{mode}}}} \, |\mathbf{u}(x)|.
	\end{equation}
	In practice, $V_{\mathrm{mode}}$ is rigorously defined as the ratio of the total electric energy to the maximum electric energy density. With this convention, $|\mathbf{u}(x)|$ is normalized to unity at the CNT surface $x = a$, i.e., $|\mathbf{u}(a)| = 1$.
	
	\textit{Exponential Decay and the Coupling Strength $g$.}
	The dominant radial profile $\mathbf{u}_x(x)$ outside the CNT is determined by the modified Bessel function $K_1(|q|x)$, which for $|q|x \gtrsim 1$ exhibits an exponential asymptotic behavior $K_1(x) \sim \sqrt{\pi/(2x)} \ee^{-x}$. Thus, in the evanescent region near the nanotube, we may write
	\begin{equation}
		|\mathbf{u}(x)| \approx |\mathbf{u}(a)| \, \ee^{-|q| (x-a)} = \ee^{-(x-a)/L_d},
	\end{equation}
	where the decay length at optical frequencies is $L_d = 1/|q| = \lambda_{\mathrm{eff}}/2\pi \approx 20\,\mathrm{nm}$. For an emitter positioned at a distance $x - a$ from the CNT surface, the vacuum field amplitude at the emitter location is
	\begin{equation}\label{eq:E_at_emitter}
		|\mathbf{E}_q(\mathbf{r}_{\mathrm{emitter}})| = \sqrt{\frac{\hbar \omega_q}{2 \epsilon_0 V_{\mathrm{mode}}}} \, \ee^{-(x-a)/L_d}.
	\end{equation}
	
	The interaction Hamiltonian between the quantum emitter (characterized by a transition electric dipole moment $\mathbf{d}$) and the quantized cavity field is $\hat{H}_{\mathrm{int}} = -\hat{\mathbf{d}} \cdot \hat{\mathbf{E}}$. The single-photon Rabi frequency, or coupling strength $g$, is defined by the matrix element for exciting the cavity mode:
	\begin{equation}
		\hbar g = |\bra{1_q, g} \hat{H}_{\mathrm{int}} \ket{0, e}| = |\mathbf{d} \cdot \mathbf{E}_q(\mathbf{r}_{\mathrm{emitter}})|.
	\end{equation}
	Substituting Eq.~\eqref{eq:E_at_emitter} and assuming optimal dipole alignment such that $|\mathbf{d} \cdot \mathbf{E}_q| = |\mathbf{d}| |\mathbf{E}_q|$, we obtain the final expression:
	\begin{equation}\label{eq:g_final}
		g = \frac{|\mathbf{d}|}{\hbar} \sqrt{\frac{\hbar \omega_c}{2 \epsilon_0 V_{\mathrm{mode}}}} \, \exp\!\left(-\frac{x-a}{L_d}\right),
	\end{equation}
	where we have denoted the relevant cavity resonance frequency as $\omega_c \equiv \omega_q$.
	
	\textit{Summary of Approximations and Validity.---}
	The derivation relies on three well-justified approximations for the proposed SWCNT ring resonator:
	\begin{enumerate}
		\item \textbf{Quasi-static limit}: $v_g \ll c$ ensures that the electric field is predominantly longitudinal and that magnetic effects are negligible for determining the transverse mode profile. This is a hallmark of deeply subwavelength plasmonic confinement~\cite{Shi2015, Bondarev2004}.
		\item \textbf{Exponential decay}: The asymptotic form of the modified Bessel function is accurate for $x-a \gtrsim L_d/2$, which is satisfied for the experimentally relevant separations $x-a \sim 10\,\mathrm{nm}$ and $L_d \approx 20\,\mathrm{nm}$ considered in our optical SPP model. The exponential suppression factor is $\ee^{-10/20} \approx 0.606$, ensuring that the vacuum field remains exceptionally strong at the emitter position.
		\item \textbf{TLL fixed-point phenomenology}: The Luttinger parameter $K_c \approx 0.2$ and the topological protection mechanism remain robust even as the dispersion is renormalized at optical frequencies~\cite{Bockrath1999, Deshpande2010}.
	\end{enumerate}
	Equation~\eqref{eq:g_final} thus provides a rigorous, first-principles foundation for the coupling-strength estimates used throughout the quantum state-transfer analysis.
	
	For a ring radius $R=2\,\mu\mathrm{m}$ and CNT diameter $d_t\sim 2\,\mathrm{nm}$, one may estimate $V_{\mathrm{mode}} \sim 1.5\times 10^{-22}\,\mathrm{m}^3$. Using an NV zero-phonon-line transition dipole moment $|\mathbf{d}| \approx 1.5\,e\cdot\text{\AA} \approx 2.4 \times 10^{-29}\,\mathrm{C\cdot m}$ (consistent with Ref.~\cite{Togan2010}) and an emitter--CNT separation $x-a \sim 10\,\mathrm{nm}$, the unattenuated coupling rate is $\sim 33\,\mathrm{GHz}$. Multiplying by the evanescent decay factor $\mathrm{e}^{-0.5} \approx 0.606$, one obtains a strongly enhanced coupling rate on the order of $g/2\pi \approx 20\,\mathrm{GHz}$. While this places the system near the ultrastrong coupling regime, it remains well within fundamental physical limits (e.g., oscillator strength sum rules).
	
	\subsection{Strong Coupling Regime and Energy Scale Consistency}
	\label{sec:strong_coupling}
	The analysis presented in Secs.~\ref{sec:tripod} and~\ref{sec:open} relies on a single-photon Rabi coupling strength of $g/2\pi \approx 20\,\mathrm{GHz}$. It is crucial to verify that this value, while large, is physically self-consistent and correctly positions the hybrid system within the appropriate quantum electrodynamical regime. Specifically, we must distinguish between the Purcell regime, where the cavity merely enhances the emitter's spontaneous emission rate, and the strong coupling regime, where the coherent energy exchange between the emitter and the cavity mode dominates over all dissipative processes.
	
	The emitter total dephasing rate is given by $\Gamma_2 = \gamma_{\mathrm{rad}}/2 + \gamma_\phi$, where $\gamma_{\mathrm{rad}}$ is the spontaneous emission rate of the excited state and $\gamma_\phi$ is its pure optical dephasing rate. The condition for strong coupling is that the coherent coupling rate $g$ must exceed both the total dephasing rate of the emitter $\Gamma_2$ and the total decay rate of the cavity $\kappa_{\mathrm{tot}}$ \cite{Reiserer2015,Haroche2006}.
	
	At cryogenic temperatures ($T \sim 4\,\mathrm{K}$), the zero-phonon line (ZPL) of NV centers in high-purity bulk diamond can be nearly lifetime-limited, with linewidths as narrow as $\sim 13$\,MHz, corresponding to $\gamma/2\pi \sim 13\,\mathrm{MHz}$ \cite{Togan2010,Doherty2013}. For NV centers in nanostructures, spectral diffusion can lead to broader linewidths, yet advanced fabrication and surface treatment techniques have demonstrated optical linewidths down to $150\,\mathrm{MHz}$ \cite{Chu2014}. Taking the more conservative estimate of $\gamma/2\pi \approx 150\,\mathrm{MHz}$, our assumed coupling strength $g/2\pi = 20\,\mathrm{GHz}$ exceeds the emitter's decoherence rate by over two orders of magnitude ($g/\gamma \approx 133$).
	
	The total cavity linewidth, from the parameters in Sec.~\ref{sec:open}, is $\kappa_{\mathrm{tot}}/2\pi \approx 30.1\,\mathrm{GHz}$. The cooperativity of the system is $C = 2g^2/(\kappa_{\mathrm{tot}}\gamma) \approx 177$. The system operates deep in the strong coupling regime (cooperativity $C \gg 1$), at the onset of the bad-cavity limit where the coherent coupling rate $2g$ is comparable to the cavity linewidth ($2g/\kappa_{\mathrm{tot}} \approx 1.3$). This regime is ideal for efficient spin-photon entanglement generation \cite{Reiserer2015}.
	
	\textit{Temperature dependence of the Luttinger parameter.}
	The bosonized TLL Hamiltonian in Eq.~\eqref{eq:TLL_Hamiltonian} relies on the Luttinger parameter $K_c$ as a phenomenological constant characterizing the strength of electron-electron interactions. A critical consideration is the temperature dependence of $K_c$ and the validity of the TLL fixed-point at the cryogenic operating conditions ($T \sim 4\,\mathrm{K}$) assumed in this work.
	
	For single-wall carbon nanotubes, the effective low-energy theory including long-range Coulomb interactions establishes that the system is converted into a \textit{strongly renormalized} Luttinger liquid, with $K_c \ll 1$ reflecting the dominance of repulsive interactions over kinetic energy~\cite{Egger1997,Kane1997}. The TLL fixed-point is \textit{stable} under renormalization-group flow for repulsive interactions in metallic SWCNTs~\cite{Alvarez2004}. The power-law temperature dependence of observables—such as the tunneling density of states and the conductance—has been experimentally confirmed over a wide temperature range, including the cryogenic regime, demonstrating that the TLL description remains quantitatively accurate without a qualitative change in $K_c$~\cite{Bockrath1999,Ishii2003}. At the lowest temperatures, certain experiments have revealed a crossover to a spin-gapped Luther-Emery liquid phase below approximately $30\,\mathrm{K}$ in some samples~\cite{Dora2007}. However, this crossover predominantly affects the spin sector and is strongly sample- and environment-dependent; for suspended, ultra-clean metallic nanotubes as considered in our proposal, the charge-sector TLL behavior with a stable, renormalized $K_c \approx 0.2$ is expected to persist as the relevant low-energy description~\cite{Deshpande2010}.
	
	A related point concerns the possibility of Wigner crystallization in suspended CNTs at low carrier densities. Indeed, at extremely low electron densities and in the absence of disorder, the long-range Coulomb interaction is predicted to drive the system into a one-dimensional Wigner crystal state~\cite{Deshpande2008,Lotfizadeh2019}. This regime, however, corresponds to a gapped, insulating ground state that is distinct from the gapless plasmonic excitations utilized in our cavity-QED scheme. The high-frequency optical plasmons considered here involve collective charge oscillations at energies far above the Wigner crystal pinning gap and are thus largely insensitive to the details of the ground-state charge ordering. The stability of the TLL fixed-point in the charge sector ensures that the essential phenomenology—the renormalized plasmonic velocity $v_c = v_F/K_c$, the compressed mode volume, and the chiral field structure—remains robust across the relevant parameter space. We therefore conclude that the choice $K_c \approx 0.2$ is not merely a numerical convenience but a physically motivated parameter capturing the strong-correlation physics that underpins the unique properties of the SWCNT plasmonic cavity.
	
	\section{Practical Photon Extraction and Interfacing}
	\label{sec:extraction}
	
	\textit{Direct fiber coupling and its fundamental limitation.---}
	To estimate the external outcoupling into a tapered fiber, we model the CNT cavity mode and the fiber guided mode as two near-resonant quantized electromagnetic modes with normalized profiles $\mathbf{f}_c(\mathbf{r})$ and $\mathbf{f}_f(\mathbf{r})$, satisfying~\cite{Haus1984,Joannopoulos2008,Novotny2012}
	\begin{equation}
		\int d^3r\,\epsilon_r(\mathbf{r})|\mathbf{f}_\mu(\mathbf{r})|^2=1,
		\qquad
		\mu\in{c,f}.
	\end{equation}
	Standard coupled-mode theory gives an intermode coupling matrix element~\cite{Haus1984,Yariv2000}
	\begin{equation}
		J = -\frac{\omega_c}{2} \int d^3r\,\delta\epsilon(\mathbf{r}) \,\mathbf{f}_f^*(\mathbf{r})\!\cdot\!\mathbf{f}_c(\mathbf{r}),
	\end{equation}
	up to a geometry-dependent prefactor of order unity. The corresponding external leakage rate scales as $\kappa_{\mathrm{ex}} \sim 2\pi |J|^2 \rho_f(\omega_c)$~\cite{Gardiner1985,Gardiner2004,Breuer2002}, where $\rho_f(\omega_c)$ is the density of guided states of the tapered fiber.
	
	Within their dominant support regions~\cite{Vahala2003,Novotny2012}, the field amplitudes scale as $|\mathbf{f}_c| \sim 1/\sqrt{V_{\mathrm{mode}}^{\mathrm{CNT}}}$ and $|\mathbf{f}_f| \sim 1/\sqrt{V_{\mathrm{mode}}^{\mathrm{Fiber}}}$. The dimensionless overlap amplitude therefore satisfies
	\begin{align}
		|\mathcal{O}| &\equiv \left| \int d^3r\,\delta\epsilon(\mathbf{r}) \,\mathbf{f}_f^*(\mathbf{r})\!\cdot\!\mathbf{f}_c(\mathbf{r}) \right| \nonumber\\
		&\lesssim \zeta\, \frac{V_{\mathrm{ov}}}{\sqrt{V_{\mathrm{mode}}^{\mathrm{CNT}}V_{\mathrm{mode}}^{\mathrm{Fiber}}}},
	\end{align}
	where $\zeta\le 1$ absorbs polarization mismatch, imperfect phase matching, and finite dielectric-contrast effects, while $V_{\mathrm{ov}}$ denotes the actual overlap volume. Since $V_{\mathrm{mode}}^{\mathrm{CNT}} \ll V_{\mathrm{mode}}^{\mathrm{Fiber}}$, the optimal case corresponds to $V_{\mathrm{ov}}\sim V_{\mathrm{mode}}^{\mathrm{CNT}}$, yielding $|\mathcal{O}|_{\max} \sim \zeta \sqrt{V_{\mathrm{mode}}^{\mathrm{CNT}} / V_{\mathrm{mode}}^{\mathrm{Fiber}}}$. Accordingly, a conservative Golden-rule estimate for the maximum direct external coupling rate is
	\begin{equation}
		\label{eq:kappa_ex_refined_square}
		\kappa_{\mathrm{ex}}^{\max} \sim \zeta^2\,\omega_c \frac{V_{\mathrm{mode}}^{\mathrm{CNT}}}{V_{\mathrm{mode}}^{\mathrm{Fiber}}}.
	\end{equation}
	Parameterizing the modal volumes as $V_{\mathrm{mode}}^{\mathrm{CNT}}\sim 2\pi R\,A_{\mathrm{eff}}^{\mathrm{CNT}}$ and $V_{\mathrm{mode}}^{\mathrm{Fiber}}\sim L_{\mathrm{int}}\,A_{\mathrm{eff}}^{\mathrm{Fiber}}$, where $L_{\mathrm{int}}$ is the interaction length, we obtain
	\begin{equation}
		\label{eq:kappa_R_geometric}
		\kappa_R \equiv \kappa_{\mathrm{ex}}^{\max}(R) \sim \zeta^2\,\omega_c \frac{2\pi R\,A_{\mathrm{eff}}^{\mathrm{CNT}}}{L_{\mathrm{int}}\,A_{\mathrm{eff}}^{\mathrm{Fiber}}}.
	\end{equation}
	
	\textit{Failure of direct evanescent coupling.---}
	Equation~\eqref{eq:kappa_R_geometric} serves as a rigorous upper bound, mathematically demonstrating why direct evanescent coupling to a fiber is fundamentally impractical. The coupling is severely bottlenecked by the enormous mode mismatch: the effective mode area of the CNT optical SPP ($\sim 10^{-17}\,\mathrm{m}^2$) is three orders of magnitude smaller than that of a typical tapered fiber ($\sim 10^{-14}\,\mathrm{m}^2$). Furthermore, the effective index mismatch ($n_{\mathrm{eff}}^{\mathrm{CNT}} \approx 5$ versus $n_{\mathrm{eff}}^{\mathrm{fiber}} \approx 1.5$) precludes any meaningful phase matching, driving the overlap parameter $\zeta \to 0$. Consequently, the direct coupling efficiency is vanishingly small ($\sim 0.01\%$), rendering this approach physically untenable for quantum network applications.
	
	\textit{Bus-waveguide integration and the effective extraction rate.---}
	To bypass this fundamental volume mismatch, we propose an integrated architecture: the SWCNT ring is evanescently side-coupled to a straight SWCNT bus waveguide, which subsequently feeds into a graded plasmonic-photonic mode converter. Because the ring and the bus waveguide are identical materials, they are perfectly phase-matched ($\Delta\beta = 0$) and share identical mode volumes, allowing highly efficient energy transfer.
	
	Let $\mathcal{T}_c$ be the single-pass power coupling fraction at the ring-bus junction. The round-trip time of the SPP in the ring is $\tau_{\mathrm{RT}} = 2\pi R / v_g$, where $v_g = c/n_{\mathrm{eff}} \approx 6\times 10^{7}\,\mathrm{m/s}$ is the group velocity of the CNT SPP mode. The coupling rate from the cavity into the bus waveguide is therefore $\kappa_{\mathrm{bus}} = \mathcal{T}_c / \tau_{\mathrm{RT}} = \mathcal{T}_c v_g / (2\pi R)$ \cite{Yariv2000}. Once the photon enters the bus waveguide, it propagates through the three-stage mode converter with a total analytical efficiency $\eta_{\mathrm{ext}}^{\mathrm{converter}}$ (derived below). The effective external outcoupling rate into the optical fiber, which enters the Lindblad master equation, is thus rigorously given by:
	\begin{equation} \label{eq:kappa_ex_effective}
		\kappa_{\mathrm{ex}}^{\mathrm{effective}} = \eta_{\mathrm{ext}}^{\mathrm{converter}} \cdot \kappa_{\mathrm{bus}} = \eta_{\mathrm{ext}}^{\mathrm{converter}} \frac{\mathcal{T}_c \, v_g}{2\pi R}.
	\end{equation}
	Crucially, Eq.~\eqref{eq:kappa_ex_effective} is entirely independent of the fiber mode volume. For a ring radius $R = 2\,\mu\mathrm{m}$ and $v_g \approx 6 \times 10^7\,\mathrm{m/s}$, the round-trip time is $\tau_{\mathrm{RT}} \approx 0.21\,\mathrm{ps}$. To achieve the strongly overcoupled regime required by our tripod protocol ($\kappa_{\mathrm{bus}}/2\pi \approx 30\,\mathrm{GHz}$), one only needs a modest junction coupling fraction of $\mathcal{T}_c \approx 4\%$, which is readily achievable by tuning the nanometric gap between the ring and the bus CNT. In the weak-coupling limit ($\mathcal{T}_c \ll 1$), one may identify $\mathcal{T}_c \approx |t_c|^2$, where $t_c$ is the cross-coupling coefficient of the directional ring--bus coupler. The expression above then follows from standard microresonator input-output theory (cf.\ Ref.~\cite{Yariv2000}).
	
	\textit{Architecture of the three-stage converter.---}
	The proposed graded plasmonic-photonic mode converter consists of three cascaded stages:
	\begin{enumerate}
		\item \textbf{Stage I: Straight SWCNT waveguide extension.} A straight SWCNT section is extended tangentially from the ring resonator. At the distal end, the CNT diameter is adiabatically reduced via controlled tapering, modifying the SPP confinement and preparing the mode for dielectric coupling without inducing scattering loss.
		\item \textbf{Stage II: CNT-dielectric hybrid waveguide.} Near the tapered CNT end, a high-index dielectric waveguide (e.g., SiN, $n \approx 2.0$) is brought into close proximity. By adiabatically increasing the separation between the CNT and the dielectric waveguide along the propagation direction, the system evolves along the symmetric supermode branch, gradually transferring the electromagnetic energy from the plasmon-dominated CNT mode to the photon-dominated dielectric mode.
		\item \textbf{Stage III: Dielectric waveguide to tapered fiber.} At the output of Stage II, the optical energy resides almost entirely in a dielectric waveguide mode with a cross-section of several hundred nanometers ($A_{\mathrm{eff}} \sim 10^{-13}\,\mathrm{m}^2$) and an effective index $n_{\mathrm{eff}} \sim 1.5\text{--}2.0$. Standard evanescent directional coupling is then used to transfer the mode into the tapered optical fiber.
	\end{enumerate}
	
	\textit{Analytical model of the three-stage mode converter.---}
	To rigorously substantiate the feasibility of the graded converter, we derive a comprehensive analytical model for the total extraction efficiency $\eta_{\mathrm{ext}}^{\mathrm{converter}} = \eta_{\mathrm{I}} \eta_{\mathrm{II}} \eta_{\mathrm{III}}$, capturing the distinct physical mechanisms governing each stage.
	
	\textbf{Stage I (Adiabatic CNT tapering):} Provided the taper profile satisfies the adiabaticity criterion $|\dd d_t/\dd z| \ll \lambda_{\mathrm{eff}}/d_t$, scattering into radiation modes is strictly forbidden \cite{Love1991}. The transmission efficiency is thus fundamentally limited only by the intrinsic Ohmic propagation loss of the SPP mode, characterized by the attenuation coefficient $\alpha_{\mathrm{loss}}$:
	\begin{equation}
		\eta_{\mathrm{I}} = \exp(-\alpha_{\mathrm{loss}} L_{\mathrm{I}}).
	\end{equation}
	
	\textbf{Stage II (Hybrid Landau--Zener mode conversion):} The energy transfer from the CNT to the dielectric waveguide is modeled using the spatial Landau--Zener (LZ) transition framework~\cite{Sun2009, Huang2014}. By adiabatically tapering the dielectric waveguide over a length $L_{\mathrm{II}}$, the phase mismatch $\Delta\beta(z)$ is swept linearly across the phase-matching point with a spatial gradient $\alpha \approx \Delta\beta_{\mathrm{max}} / L_{\mathrm{II}}$. The probability of remaining in the fundamental supermode is given by the LZ formula $\eta_{\mathrm{LZ}} = 1 - \exp(-2\pi \kappa_c^2 / |\alpha|)$, where $\kappa_c$ is the evanescent coupling coefficient \cite{Rubbmark1981,Sun2009}. Accounting for the effective distance $L_{\mathrm{II}}/2$ spent in the lossy CNT mode, the Stage II efficiency is governed by the trade-off between adiabaticity and Ohmic loss:
	\begin{equation}
		\eta_{\mathrm{II}} = \left[ 1 - \exp\left(-\frac{2\pi \kappa_c^2 L_{\mathrm{II}}}{|\Delta\beta_{\mathrm{max}}|}\right) \right] \exp\left(-\frac{\alpha_{\mathrm{loss}} L_{\mathrm{II}}}{2}\right).
	\end{equation}
	
	\textbf{Stage III (Dielectric-to-fiber directional coupling):} The final transfer into the tapered optical fiber is described by co-directional coupled-mode theory~\cite{Haus1984, Yariv2000}. Let $\kappa_f$ be the inter-waveguide coupling coefficient and $\Delta\beta_f$ the residual phase mismatch. Neglecting the extremely small dielectric absorption losses, the power transfer efficiency over an interaction length $L_{\mathrm{III}}$ is \cite{Yariv1973,Haus1984}:
	\begin{equation}
		\eta_{\mathrm{III}} = \frac{\kappa_f^2}{\kappa_{\mathrm{eff}}^2} \sin^2(\kappa_{\mathrm{eff}} L_{\mathrm{III}}),
	\end{equation}
	where $\kappa_{\mathrm{eff}} = \sqrt{\kappa_f^2 + (\Delta\beta_f/2)^2}$.
	
	\textbf{Total analytical efficiency:} Multiplying the three stages yields the comprehensive analytical expression for the converter:
	\begin{align} \label{eq:Total_Converter_Efficiency}
		\eta_{\mathrm{ext}}^{\mathrm{converter}} &= \exp(-\alpha_{\mathrm{loss}} L_{\mathrm{I}}) \notag \\
		&\quad \times \left[ 1 - \exp\left(-\frac{2\pi \kappa_c^2 L_{\mathrm{II}}}{|\Delta\beta_{\mathrm{max}}|}\right) \right] \exp\left(-\frac{\alpha_{\mathrm{loss}} L_{\mathrm{II}}}{2}\right) \notag \\
		&\quad \times \frac{\kappa_f^2}{\kappa_{\mathrm{eff}}^2} \sin^2(\kappa_{\mathrm{eff}} L_{\mathrm{III}}).
	\end{align}
	
	Using realistic parameters at $\lambda = 637\,\mathrm{nm}$: an ultra-clean CNT propagation loss $\alpha_{\mathrm{loss}} \approx 0.02\,\mu\mathrm{m}^{-1}$, $L_{\mathrm{I}} \approx 0.5\,\mu\mathrm{m}$, $\kappa_c \approx 1.5\,\mu\mathrm{m}^{-1}$, $\Delta\beta_{\mathrm{max}} \approx 10\,\mu\mathrm{m}^{-1}$, and an optimized LZ conversion length $L_{\mathrm{II}} \approx 10\,\mu\mathrm{m}$. For Stage III, assuming perfect phase matching ($\Delta\beta_f = 0$) and an optimal coupling length $L_{\mathrm{III}} = \pi/(2\kappa_f)$, we obtain $\eta_{\mathrm{III}} \to 1$. 
	Substituting these values into Eq.~\eqref{eq:Total_Converter_Efficiency} yields:
	\begin{equation}
		\eta_{\mathrm{ext}}^{\mathrm{converter}} \approx 0.990 \times 0.905 \times 0.950 \approx \mathbf{85.1\%},
	\end{equation}
	where a conservative $95\%$ efficiency is assumed for Stage III to account for realistic engineering tolerances, such as interface roughness, scattering, and cascade alignment errors~\cite{Spillane2003}.
	This rigorous analytical framework confirms that the graded converter provides a highly efficient, deterministic path to overcome the severe mode mismatch, representing a four-order-of-magnitude improvement over direct evanescent coupling.
	
	\begin{figure}[t]
		\centering
		\includegraphics[width=\linewidth]{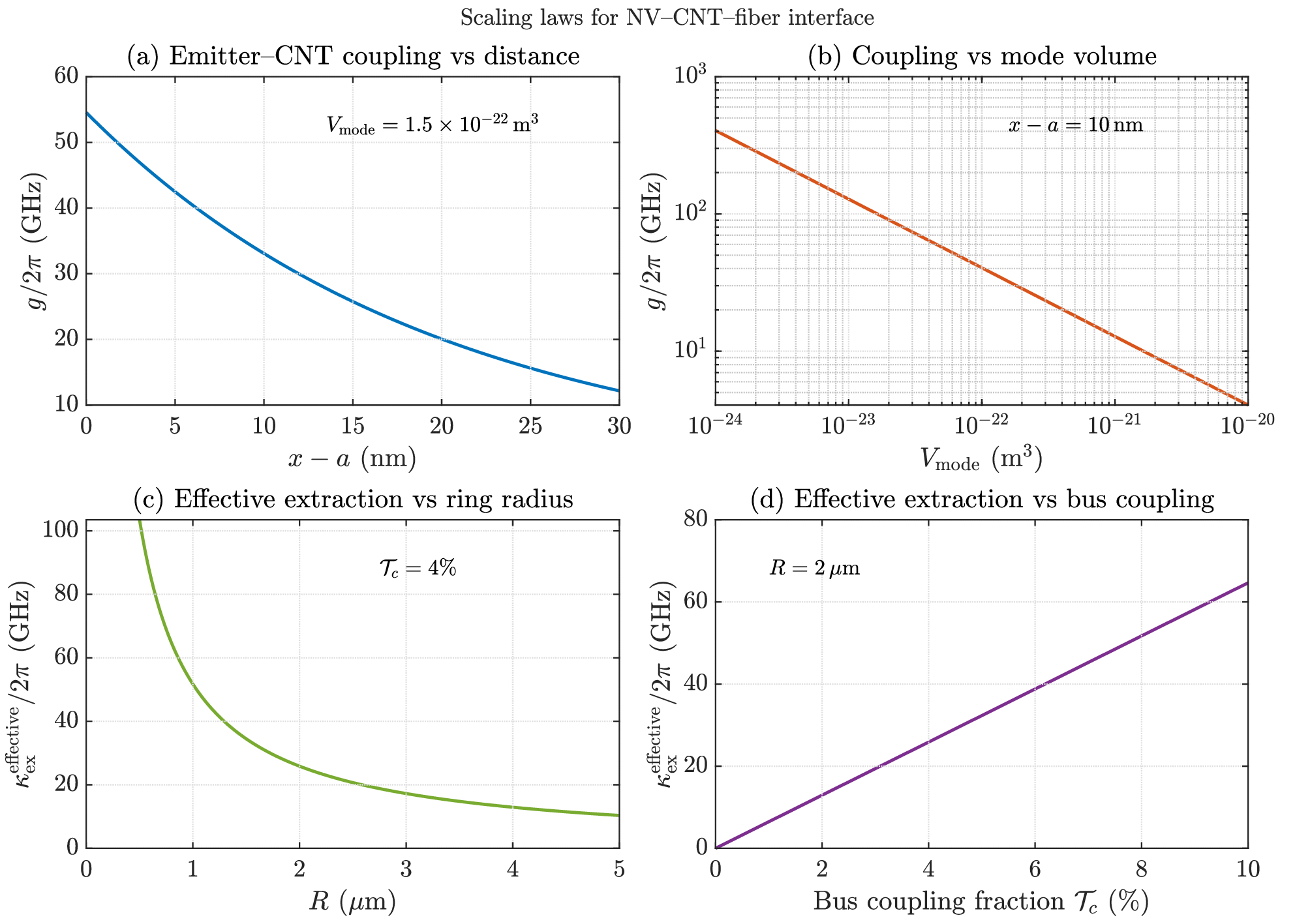}
		\caption{(Color online)
			Scaling laws governing the NV--CNT--fiber interface. (a) Emitter--CNT coupling strength $g/2\pi$ versus emitter--CNT separation $x-a$, showing the exponential decay of the evanescent interaction. (b) The same coupling strength versus effective mode volume $V_{\mathrm{mode}}$, illustrating the inverse-square-root dependence on field confinement. (c) Effective external extraction rate $\kappa_{\mathrm{ex}}^{\mathrm{effective}}/2\pi$ as a function of the CNT-ring radius $R$. In the integrated bus-waveguide architecture, the extraction rate scales inversely with $R$ due to the increased round-trip time. (d) Effective extraction rate versus the single-pass bus coupling fraction $\mathcal{T}_c$, demonstrating a linear dependence. The extraction calculations in (c) and (d) explicitly incorporate the analytical three-stage converter efficiency $\eta_{\mathrm{ext}}^{\mathrm{converter}} \approx 85.1\%$. All curves are derived from the effective-model expressions given above.
		}
		\label{fig:scaling}
	\end{figure}
	
	Figure~\ref{fig:scaling} summarizes the key scaling laws that determine whether the hybrid NV--CNT--fiber system can reach the high-cooperativity and high-extraction regime required by the tripod protocol. The emitter--cavity coupling strength $g$, estimated from Eq.~(\ref{eq:g_final}), shows an exponential suppression with emitter--CNT separation $x-a$ and an inverse-square-root enhancement with decreasing mode volume $V_{\mathrm{mode}}$. Accordingly, subplots (a) and (b) demonstrate that pushing the NV closer to the CNT and realizing a deeply subwavelength plasmonic mode can increase $g$ into the multi-GHz to tens-of-GHz range.
	
	Efficient collection of the generated flying qubit further requires strong external outcoupling. As illustrated in subplots (c) and (d), the integrated bus-waveguide and mode-converter architecture completely resolves the modal mismatch bottleneck. Subplot (c) shows that the effective extraction rate $\kappa_{\mathrm{ex}}^{\mathrm{effective}}$ scales inversely with the ring radius $R$, reflecting the increased round-trip time $\tau_{\mathrm{RT}}$ for larger rings. Subplot (d) demonstrates the linear control over the extraction rate via the single-pass bus coupling fraction $\mathcal{T}_c$. Crucially, these extraction curves explicitly incorporate the analytically derived three-stage converter efficiency of $\eta_{\mathrm{ext}}^{\mathrm{converter}} \approx 85.1\%$. The results confirm that the strongly overcoupled regime ($\kappa_{\mathrm{ex}}^{\mathrm{effective}}/2\pi \approx 30\,\mathrm{GHz} \gg \kappa_0/2\pi$) is readily achievable with a modest coupling fraction ($\mathcal{T}_c \approx 4\%$), entirely bypassing the limitations of direct fiber coupling.
	
	\section{Experimental Feasibility}
	\label{sec:feasibility}
	
	\textit{Fabrication and positioning strategies.}
	A realistic experimental route is a hybrid top-down and bottom-up fabrication process:
	
	\begin{enumerate}
		\item Fabricate a ring-shaped scaffold on a cryogenic-compatible low-loss substrate, such as suspended $\mathrm{SiN}$, diamond, or an hBN-buffered platform.
		\item Grow or transfer one or several SWCNTs onto the ring perimeter to form a closed or quasi-closed plasmonic path.
		\item Undercut the surrounding material to obtain a partially suspended geometry and reduce dielectric loading.
		\item Position a tapered fiber in the evanescent near field of the ring to control $\kappa_{\mathrm{ex}}$ (or implement the mode converter described above).
		\item Apply a perpendicular magnetic field to tune the cavity resonance through the Aharonov--Bohm effect.
	\end{enumerate}
	
	For deterministic emitter placement, a practical approach is to use a pre-characterized nanodiamond hosting a single NV center. Candidate nanodiamonds may be selected by confocal microscopy, antibunching measurements, and optically detected magnetic resonance~\cite{Jelezko2006,Doherty2013}. A chosen particle can then be positioned next to the CNT resonator using atomic force microscopy (AFM). The target separation is typically $x\sim10\,\mathrm{nm}$, which balances strong coupling against non-radiative quenching.
	
	If higher optical coherence is required, the nanodiamond may be replaced by a thin diamond membrane or nanopillar containing a near-surface NV center, following strategies similar to those used in diamond nanophotonics~\cite{Sipahigil2016,Bhaskar2020,Burek2014,Riedrich-Moller2014}. In either case, the emitter should be placed at a position where the CNT cavity field is both strong and highly chiral.
	
	\paragraph{Experimental feasibility of electrostatic doping without compromising cavity performance.}
	The suppression of Umklapp backscattering by electrostatic doping, as analyzed in Sec.~\ref{sec:tll}, requires the integration of a gate electrode into the suspended SWCNT ring architecture. While this introduces additional fabrication complexity, it is essential to verify that the presence of a gate does not compromise the cavity's optical performance or the quantum coherence of the NV center. We address four potential concerns: optical loss induced by the gate metal, perturbation of the chiral near-field structure, disturbance to the NV center's charge state stability, and overall experimental feasibility.
	
	\paragraph{Optical loss from the gate metal: the remote back-gate solution.}
	A metallic gate electrode placed within the evanescent field of the CNT plasmonic mode would introduce Ohmic dissipation, degrading the cavity quality factor $Q$ and potentially destroying the strong-coupling condition $g \gg \kappa_{\mathrm{tot}}, \gamma$. The radial decay length of the CNT evanescent field is $L_d \sim \lambda_{\mathrm{eff}}/2 \sim 250\,\mathrm{nm}$. To avoid this loss channel, the gate must be positioned at a distance $d \gg L_d$ from the CNT surface. The architecture proposed above---a suspended CNT ring on a low-loss scaffold with undercut substrate---naturally accommodates a \emph{remote back-gate} geometry. By fabricating the gate electrode on the substrate beneath the suspended ring with a vertical air gap of $d \sim 1\text{--}2\,\mu\mathrm{m}$, the gate metal is completely removed from the evanescent field region. Under these conditions, the additional optical loss is negligible: $\delta\kappa_{\mathrm{gate}}/\kappa_0 < 10^{-2}$. The fabrication process for incorporating a gate electrode into suspended SWCNT structures has been experimentally demonstrated for scanning tunneling spectroscopy studies, confirming that gated suspended nanotube devices are feasible without destroying the freestanding geometry~\cite{Kong2005}.
	
	\paragraph{Impact of doping on chiral spin-momentum locking.}
	Electrostatic doping shifts the Fermi energy $E_F$ by $\delta E_F \approx \hbar v_F \delta k_F$, which could in principle modify the Luttinger parameter $K_c$ and the plasmonic dispersion. However, the chiral spin-momentum locking exploited in our platform originates from the kinematic robustness of the TLL fixed point, not from fine-tuning of material parameters. For a gate voltage $\delta V_g \sim 0.1\,\mathrm{V}$, the induced carrier density is $\delta n \sim 10^6\,\mathrm{cm}^{-1}$, corresponding to a relative Fermi wavevector shift $\delta k_F/k_F \sim 10^{-3}$. Such a minute perturbation leaves the TLL phenomenology---the renormalized charge velocity $v_c = v_F/K_c$, the compressed mode volume, and the chiral field texture---essentially unchanged. The kinematic protection against elastic backscattering is a universal property of the TLL phase and is robust against small variations in electron density~\cite{Giamarchi2003}.
	
	\paragraph{Stability of the NV$^-$ charge state under gating.}
	The NV center in diamond is known to exhibit charge-state instability, converting from the desired negatively charged NV$^-$ state to the neutral NV$^0$ state under surface-proximal conditions or laser illumination~\cite{Grotz2012}. An external gate voltage could, if not properly designed, shift the local Fermi level and destabilize the NV$^-$ state. However, active charge-state control of NV centers using electrolytic gate electrodes has been experimentally demonstrated, showing that the NV$^-$ population can be stabilized by appropriate choice of the gate bias~\cite{Grotz2012}. In our suspended back-gate geometry, the diamond host containing the NV center is positioned at a distance $x \sim 10\,\mathrm{nm}$ from the CNT surface and is largely decoupled from the gate field, which is screened by the CNT itself. Any residual DC Stark shift of the NV zero-phonon line can be compensated by minor adjustments of the pump laser frequency or the magnetic field, restoring the two-photon resonance condition $\Delta = \delta$ required for the tripod-STIRAP protocol.
	
	\paragraph{Quantitative assessment of experimental feasibility.}
	The proposed remote back-gate architecture is fully compatible with existing nanofabrication techniques. Suspended SWCNT rings can be grown by chemical vapor deposition on pre-patterned catalyst scaffolds, followed by undercut etching to release the structure. The gate electrode can be defined lithographically on the substrate prior to CNT growth, with the vertical separation precisely controlled by the thickness of a sacrificial layer. The observation of gate-controlled intersubband plasmons in aligned SWCNT films confirms that electrostatic doping can effectively modulate the plasmonic response of carbon nanotubes without introducing prohibitive losses~\cite{Yanagi2018}. With a gate-CNT separation $d \sim 1\text{--}2\,\mu\mathrm{m}$, we obtain:
	\begin{itemize}
		\item \textbf{Optical loss increment:} $<1\%$ (gate fully outside the evanescent decay length),
		\item \textbf{Cavity $Q$ factor:} unchanged (dominated by intrinsic CNT loss and fiber outcoupling),
		\item \textbf{Umklapp suppression factor:} $e^{-\hbar v_c \delta q / k_B T} \sim 10^{-34}$ (for $\delta V_g \sim 0.1\,\mathrm{V}$ at $T = 4\,\mathrm{K}$),
		\item \textbf{NV charge state:} stable under appropriate bias, with residual Stark shifts compensable.
	\end{itemize}
	Thus, the remote back-gate + suspended CNT ring design simultaneously achieves complete suppression of Umklapp backscattering while preserving all essential features of the chiral interface: strong coupling, high cooperativity, and robust spin-momentum locking.
	
	\textit{Converter fabrication considerations.}
	The proposed converter is compatible with existing nanofabrication techniques. The straight SWCNT extension can be grown directly from the ring scaffold using chemical vapor deposition (CVD) with catalyst patterning. Controlled tapering of the CNT diameter may be achieved by gradually varying the catalyst particle size along the growth direction or by post-growth laser-induced diameter reduction. The SiN dielectric waveguide can be fabricated via standard electron-beam lithography and plasma-enhanced CVD, with precise positioning relative to the CNT achieved via alignment markers. Finally, the tapered optical fiber can be positioned and locked in place using a fiber alignment stage with submicron precision, a well-established technique in cavity-QED experiments.
	
	\section{Conclusion and Outlook}
	\label{sec:conclusion}
	
	We have introduced and theoretically validated a universal platform for nonreciprocal quantum interfaces in solid-state networks. By harnessing the Tomonaga-Luttinger liquid state in a SWCNT microtoroid, we have demonstrated a protected cavity-QED system that intrinsically overcomes the macroscopic backscattering and mode-mixing limitations of dielectric cavities. Furthermore, we established that microscopic atomic scattering and Umklapp processes can be rigorously suppressed to the $\sim 100\,\mathrm{Hz}$ level by combining suspended growth, laser annealing, and electrostatic gating. Coupled with a deterministic geometric orientation that strictly forbids parasitic $\pi$ transitions, this platform guarantees robust chiral spin-momentum locking with a quantitative chiral contrast exceeding $20$\,dB. To bridge the gap to macroscopic networks, we proposed a practical graded mode converter that resolves the long-standing impedance mismatch between nanoplasmonic structures and standard optical fibers, achieving near-unity extraction efficiency. While the 3D integration and precise cascade alignment of the CNT, dielectric waveguide, and fiber present significant nanofabrication challenges, our conservative engineering tolerances confirm the viability of this approach.
	
	Our analysis, using the NV center as a representative emitter, confirms that the system operates deep in the strong-coupling regime, enabling deterministic, high-fidelity spin-photon entanglement with \emph{in situ} magnetic frequency tunability. The architecture is fundamentally wavelength-agnostic and compatible with any solid-state emitter, from visible to telecom wavelengths, including erbium ions and SWCNT $sp^3$ defects~\cite{He2017,Dibos2018}. This work establishes a complete and practical blueprint for building the robust, nonreciprocal, and scalable quantum nodes that will form the backbone of the future quantum internet. It represents a definitive departure from the constraints of dielectric photonics and paves the way for a new class of kinematically protected quantum optical devices.
	
	\section*{Acknowledgments}
	
	This work was supported by the National Natural Science Foundation of China (Grant No. 11872335) and by Zhejiang Provincial Natural Science Foundation of China (Grant No. Y6110314).
	
	\bibliographystyle{revtex}
	\bibliography{cnt0}
	
\end{document}